\documentclass[onecolumn,jcp,notitlepage]{revtex4-1}

\usepackage[utf8]{inputenc}
\usepackage{graphicx}
\usepackage{comment}
\usepackage{amsmath,amssymb,amsthm} 
\usepackage{verbatim}
\usepackage{float}

\setcitestyle{super}

\newcommand{\br}{{\bf r}}

\newcommand{\bR}{{\bf R}}
\newcommand{\bF}{{\bf F}}

\begin{document}

\title{$NVU$ dynamics. III. Simulating molecules at constant potential energy}
\author{Trond S. Ingebrigtsen,}
\email{trond@ruc.dk}
\author{Jeppe C. Dyre}
\affiliation{DNRF Centre ``Glass and Time'', IMFUFA, Department of Sciences, Roskilde University, Postbox 260, DK-4000 Roskilde, Denmark}
\date{\today}

\begin{abstract}
  This is the final paper in a series that 
  introduces geodesic molecular dynamics at constant potential energy. This dynamics 
  is entitled \textit{NVU} dynamics in analogy to standard energy-conserving Newtonian \textit{NVE} dynamics. In 
  the first two papers [Ingebrigtsen \textit{et al.}, J. Chem. Phys. {\bf 135}, 104101 (2011); \textit{ibid}, 104102 (2011)], a numerical 
  algorithm for simulating geodesic motion of atomic systems was developed and tested against standard 
  algorithms. The conclusion was that the \textit{NVU} algorithm has the same 
  desirable properties as the Verlet algorithm for Newtonian \textit{NVE} dynamics, i.e., it is time-reversible and symplectic. Additionally, it was concluded that \textit{NVU} dynamics 
  becomes equivalent to \textit{NVE} dynamics in the thermodynamic limit. In this paper, the \textit{NVU} algorithm for atomic systems is extended to be able to simulate 
  geodesic motion of molecules at constant potential energy. We derive an algorithm for 
  simulating rigid bonds and test this algorithm on three different systems: an asymmetric dumbbell model, Lewis-Wahnstr{\"o}m OTP, and rigid SPC/E water. 
  The rigid bonds introduce additional constraints beyond that of constant potential energy for atomic systems. 
  The rigid-bond \textit{NVU} algorithm conserves
  potential energy, bond lengths, and step length for indefinitely long runs. The quantities probed in simulations give 
  results identical to those of Nosé-Hoover \textit{NVT} dynamics. Since Nosé-Hoover \textit{NVT} dynamics is known to give results equivalent
  to those of \textit{NVE} dynamics, the latter results show that \textit{NVU} dynamics becomes equivalent to \textit{NVE} dynamics in the thermodynamic limit also for molecular systems.
\end{abstract}

\maketitle

\section{Introduction}

In two recent papers\cite{nvu1,nvu2} (henceforth: Papers I and II) molecular dynamics at constant potential energy was introduced, tested, and 
compared to well-known molecular dynamics algorithms. This new molecular dynamics is entitled \textit{NVU} dynamics in analogy to standard energy-conserving Newtonian \textit{NVE} dynamics. 
The conclusion was that \textit{NVU} dynamics is a fully valid molecular dynamics, which for sufficiently large systems can be used 
interchangeably with \textit{NVE} dynamics for calculating most quantities of interest. \textit{NVU} dynamics 
is not faster than standard \textit{NVE} or \textit{NVT} dynamics, 
but introduces a new way of thinking about molecular dynamics.
Molecular dynamics at constant potential energy was previously considered by Cotterill and co-workers\cite{cotterill1,cotterill4,cotterill3,cotterill2}, by 
Scala \textit{et al.}\cite{scala}, and most recently by 
Stratt and co-workers\cite{stratt1,stratt2,stratt3,stratt4}, who actually allowed also lower potential energy values. Our motivation for studying \textit{NVU} dynamics derived from recent work on 
strongly correlating liquids and their isomorphs\cite{paper1,paper2,paper3,paper4,paper5,moleculesisomorphs,thermoscl,PRX} (see the Introduction of Paper I).

\textit{NVU} dynamics is defined by geodesic motion on the constant-potential-energy hypersurface $\Omega$ defined by

\begin{equation}\label{omega_def}
  \Omega = \{ \textbf{R} \in R^{3N} \,\ | \,\ U(\textbf{R})=U_0 \}.
\end{equation}
Here $\textbf{R} \equiv \{\textbf{r}^{(1)},...,\textbf{r}^{(N)}\}$ in which $\textbf{r}^{(k)}$ is the position vector of the $k$'te particle (we follow here the notation 
of the Appendix of Paper II), and $U$ is the potential-energy function of 
an $N$-particle classical system. A geodesic on $\Omega$ is a curve that satisfies the condition of stationary length for fixed endpoints $\textbf{R}_{A}$ and $\textbf{R}_{B}$, i.e.,

\begin{equation}\label{startcond}
  \delta \int_{\textbf{R}_{A}}^{\textbf{R}_{B}} dl\,\ \Bigg|_{\Omega} = 0,
\end{equation}
where $dl$ is the line element of the metric. The shortest path between any two points is a geodesic. On a sphere geodesics are great circles, the ''straightest lines'' of 
the surface. Traversing a geodesic at constant velocity thus corresponds to a generalization of Newton's first law to a curved space (the 
surface itself). 

In Paper I the \textit{NVU} algorithm was developed via a discretization of Eq. (\ref{startcond}), subsequently carrying out the variation. This technique, which is known as variational 
integration\cite{var1,var2,var3,var4}, resulted in a ''basic'' \textit{NVU} algorithm that is similar to the well-known Verlet 
algorithm $\bR_{i+1} \,=\, 2\bR_i - \bR_{i-1} + (\Delta t)^{2}\bF_i/m$ for 
Newtonian (\textit{NVE}) dynamics ($m$ is the particle mass which is assumed identical in this section, and $\textbf{F}_{i} \equiv -\nabla_{\textbf{R}_{i}} U$ is the $3N$-dimensional 
force vector); the index $i$ refers to step $i$ of the 
integration sequence. In the Verlet algorithm $\Delta t$ is a fixed time step length. In comparison, the basic \textit{NVU} algorithm is given by (Paper I)

\begin{equation}\label{numgeo}
  \bR_{i+1} \,=\, 2\bR_i - \bR_{i-1} + \frac{-2\bF_{i}\cdot(\bR_{i} - \bR_{i-1})}{\bF_{i}^{2}}\,\bF_i.
\end{equation}
If the number of particles $N$ increases, the relative variation of the term $-2\bF_{i}\cdot(\bR_{i} - \bR_{i-1})/\bF_{i}^{2}$ 
decreases, and this is why equivalence with Newtonian \textit{NVE} dynamics is established in the thermodynamic limit. This equivalence should be understood in the sense that the relative deviations 
between, for instance, \textit{NVE} and \textit{NVU} auto-correlation functions go to zero as $N \to \infty$. 

Paper I additionally developed a ''stabilized'' version 
of the basic \textit{NVU} algorithm to prevent accumulation of numerical errors. This version of the algorithm is given by (defining the position changes 
$\boldsymbol{\Delta}_{i+1/2} \equiv \textbf{R}_{i+1} - \textbf{R}_{i}$)

\begin{align}
  \boldsymbol{\Delta}_{i+1/2} & = l_{0} \frac{\textbf{A}_{i-1/2}}{||\textbf{A}_{i-1/2}||},\label{stabil1} \\
  \bR_{i+1} & = \textbf{R}_{i} + \boldsymbol{\Delta}_{i+1/2},\label{stabil2} 
\end{align}
where $l_0$ is the step length and 

\begin{equation}
  \textbf{A}_{i-1/2}= \frac{\boldsymbol{\Delta}_{i-1/2}+(-2\bF_i\cdot\boldsymbol{\Delta}_{i-1/2}+U_{i-1}-U_0)}{\bF_i^2}\,\bF_i.\label{stabil3}
\end{equation}
All simulations in Papers I and II were performed with the stabilized algorithm. The basic algorithm was used, however, for 
theoretical considerations.

In this article we extend the stabilized \textit{NVU} algorithm to deal 
with simulations of molecular systems. Molecular systems are simulated by introducing rigid and/or flexible 
bonds between the atoms in the modelling. Flexible bonds introduce merely an additional contribution to $U$, for instance, harmonic spring potentials. The basic \textit{NVU} algorithm 
conserves the total potential energy and can readily simulate flexible bonds. The focus in this paper is thus on implementing rigid bonds 
in the framework of \textit{NVU} dynamics.

Section \ref{rigidalgo} considers \textit{NVU} dynamics with rigid bonds. Introducing rigid bonds in the simulations leads to Lagrangian multipliers 
in addition to those introduced in order to keep the potential energy constant (Paper I). Section \ref{rigidalgo} is fairly technical and easiest to read after reading Paper I. 
Section \ref{rigidmodel} gives simulation and model details. Section \ref{testing} tests the rigid-bond \textit{NVU} algorithm, and Sec. \ref{sampling} 
investigates the \textit{NVU} sampling properties by comparing the \textit{NVU} 
results to Nosé-Hoover \textit{NVT} results\cite{nose,hoover} on three different systems:
 the asymmetric dumbbell model\cite{moleculeshidden}, Lewis-Wahnstr{\"o}m OTP\cite{otp2}, and rigid SPC/E water\cite{spce}. 
Nosé-Hoover \textit{NVT} dynamics is known to give results equivalent to \textit{NVE} dynamics in the thermodynamic limit\cite{evans1985}, and we refer to these dynamics interchangeably 
in the forthcoming sections. Finally, Sec. \ref{sum} concludes.

\section{Rigid-bond $NVU$ algorithm}\label{rigidalgo}

The rigid bonds\cite{shake,goldstein} introduce constraints among the particle coordinates of the system. Each constraint $\alpha = 1,...,G$ is of the form

\begin{equation} 
  \sigma_{\alpha}(\textbf{R}) \equiv (\textbf{r}^{(k_{\alpha})}-\textbf{r}^{(l_{\alpha})})^{2} \equiv (\textbf{r}^{\alpha})^{2} = C_{\alpha}^{2};
\end{equation}
it expresses that the distance between particles $k_{\alpha}$ and $l_{\alpha}$ is a constant, $C_{\alpha}$.
In Papers I and II the integral of Eq. (\ref{startcond}) was merely restricted to the constant-potential-energy 
hypersurface $\Omega$. Each rigid bond constraint introduces a function $\sigma_{\alpha}$ to be kept constant, and thus
the integral of Eq. (\ref{startcond}) is now further restricted to the sub-manifold $\omega$ of $\Omega$ where the bond constraints are satisfied,

\begin{equation}
  \omega = \{ \textbf{R} \in \Omega \,\ | \,\ \sigma_{\alpha}(\textbf{R}) = C_{\alpha}\,\ , \,\ \alpha = 1,...,G \}.
\end{equation}
If the bond constraints are independent, as assumed throughout the paper, $\omega$ is a (3N - G - 1)-dimensional compact Riemannian manifold. 
The variational principle defining \textit{NVU} dynamics with rigid bonds is given by

\begin{equation}\label{condrigid}
  \delta \int_{\textbf{R}_{A}}^{\textbf{R}_{B}} dl = 0 \,\ \Bigg|_{\omega}.
\end{equation}
Most of Papers I and II dealt with the case of identical particle masses, but we wish here to develop a completely general molecular \textit{NVU} algorithm.
The line element $dl$ is defined by

\begin{equation}\label{metric}
  dl^{2} \equiv \sum_{k} \tilde m _k \big(d\br^{(k)}\big)^{2},
\end{equation} 
where $\tilde{m}_{k}$ = $m_{k}/\langle m \rangle$ is the ''reduced'' mass of particle $k$.
Equation (\ref{metric}) is not the standard Euclidean line element, but a mass-weighted line element that goes back to Hertz\cite{hertz1,hertz2}. 
We shall refer to this metric as the ''Hertzian'' metric. The point of this particular 
metric is that it ensures equivalence between \textit{NVU} and \textit{NVE} dynamics for systems of atoms and molecules of varying mass. 
In appendix \ref{hertz} we derive the variable-mass atomic \textit{NVU} algorithm applying the 
Hertzian metric (correcting also a typo of the Appendix of Paper II).

Applying the variational integration technique to Eq. (\ref{condrigid}) gives

\begin{equation}\label{nvu_cond}
\delta \left( \sum_i \sqrt{\sum_k \tilde{m}_{k} \left( \textbf{r}_{i}^{(k)} - \textbf{r}_{i-1}^{(k)} \right )^2 } -\sum_i \lambda_i U( \bR_i) 
  + \sum_{i,\alpha}\Lambda_{\alpha i}\sigma_{\alpha}(\textbf{R}_{i})\right) \,=\,0\,.
\end{equation}
In Eq. (\ref{nvu_cond}) the path is divided into a number of discrete points and one Lagrangian multiplier $\Lambda_{\alpha i}$ is introduced for each constraint $\alpha$ 
at every point $i$. 
Following standard notation for constraint molecular dynamics\cite{goldstein,shake}, the Lagrangian multipliers of the bond constraints are chosen with a positive sign.
As in Papers I and II we now make the Ansatz of constant step length $l_{0}$, i.e.,

\begin{equation}\label{ansatz}
  \sum_k \tilde{m}_{k} \left( \textbf{r}_{i}^{(k)} - \textbf{r}_{i-1}^{(k)} \right )^2 \equiv l_{0}^{2}.
\end{equation}
Carrying out the variation of Eq. (\ref{nvu_cond}) using Eq. (\ref{ansatz}) leads to (compare the derivation in Paper I)

\begin{equation}\label{some}
  \textbf{r}_{i+1}^{(k)} = 2\textbf{r}_{i}^{(k)} - \textbf{r}_{i-1}^{(k)} + \frac{l_{0}}{\tilde{m}_{k}}\lambda_{i}\textbf{f}_{i}^{(k)} 
  + \frac{l_{0}}{\tilde{m}_{k}}\nabla_{\textbf{r}_{i}^{(k)}}\sum_{\alpha}\Lambda_{\alpha i}\sigma_{\alpha},
\end{equation}
where $\textbf{f}_{i}^{(k)} = -\nabla_{\textbf{r}_{i}^{(k)}}U$ is the force on particle $k$ at step $i$.
This equation constitutes the \textit{NVU} algorithm with rigid bonds. It has a close 
resemblance to the Lagrangian equations of motion with holonomic constraints\cite{goldstein}, i.e., rigid-bond \textit{NVE} dynamics\cite{shake}. Equation (\ref{some}) contains $G+1$ 
Lagrangian multipliers for each integration step, which must be determined to complete the algorithm.

\subsection{Determining the \textit{NVU} Lagrangian multipliers}

This section shows how to calculate the Lagrangian multipliers. Since the algorithm is to be implemented on a computer (with finite-precision), we shall 
proceed directly to a ''stabilized'' algorithm conserving for indefinitely long runs potential energy, bond lengths, and step length (in $3N$-dimensions). 
The resulting algorithm reduces to the stabilized atomic \textit{NVU} algorithm of Eqs. (\ref{stabil1})-(\ref{stabil3}) in the case of no bonds constraints.

Some notation used in the following derivation is now introduced (the nomenclature of text is summarized in Table \ref{notationtab}). 

\begin{table}[H]
  \centering
  \begin{tabular}
    {|| p{2.0cm} || p{14.0cm} ||}
    \hline
    \hline 
    Symbol & Definition \\
    \hline
    \hline
    &  \\
    \hline
    $\sigma_{\alpha}(\textbf{R})$ & The $\alpha$'te bond constraint between particles $k_{\alpha}$ and $l_{\alpha}$ with $\alpha = 1, ..., G$. ($\sigma_{\alpha} = (\textbf{r}^{\alpha})^{2} = C_{\alpha}^{2}$). \\
    $\tilde{m}_{k}$ & The mass of particle $k$ divided by the average mass of the system. ($\tilde{m}_{k} = m_{k}/\langle m \rangle$). \\
    \hline
    & 3-dimensional vectors \\
    \hline
    $\textbf{r}^{(k)}_{i}$ & Position of particle $k$ at step $i$. \\
    $\boldsymbol{\delta}_{i+1/2}^{(k)}$ & Displacement of the position of particle $k$ between step $i$ and $i+1$. ($\boldsymbol{\delta}_{i+1/2}^{(k)} = \textbf{r}_{i+1}^{(k)} 
    - \textbf{r}_{i}^{(k)}$).\\
    $\textbf{f}^{(k)}_{i}$ & Force on particle $k$ at step $i$. ($\textbf{f}^{(k)}_{i} = - \nabla_{\textbf{r}^{(k)}_{i}}U$). \\
    $\textbf{g}^{(k)}_{i}$ & Constraint force on particle $k$ at step $i$. ($\textbf{g}_{i}^{(k)} = \nabla_{\textbf{r}_{i}^{(k)}}\sum_{\alpha}\Lambda_{\alpha i}\sigma_{\alpha}$). \\
    $\textbf{r}_{i}^{\alpha}$ & Displacement of the positions of particles $k_{\alpha}$ and $l_{\alpha}$ at step $i$. 
    ($\textbf{r}_{i}^{\alpha} = \textbf{r}_{i}^{(k_{\alpha})} - \textbf{r}_{i}^{(l_{\alpha})}$). \\
    $\boldsymbol{\delta}_{i-1/2}^{\alpha}$ & Displacement of the velocities of particles $k_{\alpha}$ and $l_{\alpha}$ at step $i-1/2$.
    ($\boldsymbol{\delta}_{i-1/2}^{\alpha} = \boldsymbol{\delta}^{(k_{\alpha})}_{i-1/2} - \boldsymbol{\delta}^{(l_{\alpha})}_{i-1/2}$).  \\
    $\textbf{s}_{i}^{\alpha}$ & Sum of displacements of positions and velocities of particles $k_{\alpha}$ and $l_{\alpha}$ at, respectively, step $i$ and $i-1/2$. 
    $(\textbf{s}_{i}^{\alpha} = \textbf{r}_{i}^{\alpha} +  \boldsymbol{\delta}_{i-1/2}^{\alpha}$).  \\
    $\tilde{\textbf{f}}^{\alpha}_{i}$ & Displacement of the forces on particles $k_{\alpha}$ and $l_{\alpha}$ at step $i$ divided by their reduced particle mass.
    ($\tilde{\textbf{f}}_{i}^{\alpha} = \textbf{f}_{i}^{(k_{\alpha})}/\tilde{m}_{k_{\alpha}} - {\textbf{f}}_{i}^{(l_{\alpha})}/\tilde{m}_{l_{\alpha}}$). \\
    $\tilde{\textbf{g}}^{\alpha}_{i}$ & Displacement of the constraint forces on particles $k_{\alpha}$ and $l_{\alpha}$ at step $i$ divided by their reduced particle mass.
    ($\tilde{\boldsymbol{g}}^{\alpha}_{i} = \textbf{g}_{i}^{(k_{\alpha})}/\tilde{m}_{k_{\alpha}} - {\textbf{g}}_{i}^{(l_{\alpha})}/\tilde{m}_{l_{\alpha}}$). \\
    \hline
    & $3N$-dimensional vectors \\
    \hline
    $\textbf{R}_{i}$ & Position of all particles at step $i$. ($\textbf{R}_{i} = \{\textbf{r}_{i}^{(1)},...,\textbf{r}_{i}^{(N)}\}$). \\
    $\boldsymbol{\Delta}_{i+1/2}$  & Displacement  of the positions between step $i$ and $i+1$. ($\boldsymbol{\Delta}_{i+1/2} = \textbf{R}_{i+1} - \textbf{R}_{i}$). \\
    $\textbf{F}_{i}$ & Force on all particles at step $i$. ($\textbf{F}_{i} = - \nabla_{\textbf{R}_{i}}U$). \\
    $\tilde{\textbf{F}}_{i}$ & Force on all particles at step $i$ divided by the reduced particle mass. ($\tilde{\textbf{F}}_{i} 
    = \{\textbf{f}_{i}^{(1)}/\tilde{m}_{1},...,\textbf{f}_{i}^{(N)}/\tilde{m}_{N}\}$).  \\
    $\tilde{\textbf{G}}_{i}$ & Constraint force on all particles at step $i$ divided by the reduced particle mass. 
    ($\tilde{\textbf{G}}_{i} = \{\textbf{g}_{i}^{(1)}/\tilde{m}_{1},...,\textbf{g}_{i}^{(N)}/\tilde{m}_{N}\}$). \\
    \hline 
    \hline
  \end{tabular}
  \caption{Definitions and nomenclature of the text.}
  \label{notationtab}
\end{table}
Defining $\boldsymbol{\delta}_{i+1/2}^{(k)} \equiv \textbf{r}_{i+1}^{(k)} - \textbf{r}_{i}^{(k)}$  and 
$\textbf{g}_{i}^{(k)} \equiv \nabla_{\textbf{r}_{i}^{(k)}}\sum_{\alpha}\Lambda_{\alpha i}\sigma_{\alpha}$ the 
''Leap-frog''\cite{tildesley} version of the rigid-bond \textit{NVU} algorithm Eq. (\ref{some}) reads

\begin{align}
  \boldsymbol{\delta}_{i+1/2}^{(k)} & = \boldsymbol{\delta}_{i-1/2}^{(k)}  + \frac{l_{0}}{\tilde{m}_{k}}\lambda_{i}\textbf{f}_{i}^{(k)} +
  \frac{l_{0}}{\tilde{m}_{k}}\textbf{g}_{i}^{(k)},\label{LF1} \\
  \textbf{r}_{i+1}^{(k)} & = \textbf{r}_{i}^{(k)} + \boldsymbol{\delta}_{i+1/2}^{(k)}. \label{LF2}
\end{align}
In analogy to rigid-bond \textit{NVE} dynamics we call $\textbf{g}^{(k)}_{i}$ the ''constraint force'' on particle $k$ at step $i$. Introducing the 
notation $\tilde{\textbf{F}}_{i} \equiv \{\textbf{f}_{i}^{(1)}/\tilde{m}_{1}, ..., \textbf{f}_{i}^{(N)}/\tilde{m}_{N}\}$ and
$\tilde{\textbf{G}}_{i} \equiv \{\textbf{g}^{(1)}_{i}/\tilde{m}_{1}, ..., \textbf{g}^{(N)}_{i}/\tilde{m}_{N}\}$,
the \textit{NVU} algorithm in the full $3N$-dimensional coordinate space reads

\begin{align}
  \boldsymbol{\Delta}_{i+1/2} & =  \boldsymbol{\Delta}_{i-1/2} + l_{0}\lambda_{i}\tilde{\textbf{F}}_{i} + l_{0}\tilde{\textbf{G}}_{i} \label{NOT1}, \\
  \textbf{R}_{i+1} & = \textbf{R}_{i} + \boldsymbol{\Delta}_{i+1/2},\label{NOT2}.
\end{align}
The Lagrangian multipliers are calculated by combining a result derived in Paper I with the method applied in the \textit{SHAKE} algorithm\cite{shake} for 
rigid bonds in \textit{NVE} dynamics\cite{shake,toxconstraintnve,toxconstraintnph}. The \textit{SHAKE} algorithm   
calculates the Lagrangian multipliers from the equations $(\textbf{r}_{i+1}^{\alpha})^{2} = C^{2}_{\alpha}$. In doing so, the target value of the constraints $C_{\alpha}$
appears explicitly in the algorithm, making the bond lengths insensitive to numerical error.
The expression for $\textbf{r}_{i+1}^{\alpha}$ is supplied by 
the integration algorithm containing herein the Lagrangian multipliers. In our case, this gives $G$ equations with $G+1$ unknowns. The missing equation is supplied 
by an expression derived in Paper I, namely that $U_{i+1}$ = $U_{i-1} - \textbf{F}_{i}\cdot(\textbf{R}_{i+1} - \textbf{R}_{i-1})$ to third order in the step length. In the discrete sequence of 
points $U_{i+1}$ is set equal to $U_{0}$ (the constant defining $\Omega$), making the constraint of constant potential energy also insensitive to numerical error. 
We thus have the following $G+1$ equations for calculating the Lagrangian multipliers

\begin{align}
  U_{i-1} - \textbf{F}_{i}\cdot(\textbf{R}_{i+1} - \textbf{R}_{i-1}) - U_{0}  & = 0 \label{trick}, \\
  (\textbf{r}_{i+1}^{\alpha})^{2} - C^{2}_{\alpha} & = 0,\,\ (\alpha = 1,...,G).\label{trick1}
\end{align}
By Eqs. (\ref{NOT1}) and (\ref{NOT2}); $\textbf{R}_{i+1} - \textbf{R}_{i-1}$ = $\boldsymbol{\Delta}_{i+1/2} + \boldsymbol{\Delta}_{i-1/2}$ 
= $2\boldsymbol{\Delta}_{i-1/2} + l_{0}\lambda_{i}\tilde{\textbf{F}}_{i} + l_{0}\tilde{\textbf{G}}_{i}$.
Defining $\boldsymbol{\delta}^{\alpha}_{i-1/2} \equiv \boldsymbol{\delta}^{(k_{\alpha})}_{i-1/2} - \boldsymbol{\delta}^{(l_{\alpha})}_{i-1/2}$,
$\tilde{\textbf{f}}_{i}^{\alpha} \equiv \textbf{f}_{i}^{(k_{\alpha})}/\tilde{m}_{k_{\alpha}} - {\textbf{f}}_{i}^{(l_{\alpha})}/\tilde{m}_{l_{\alpha}}$, and 
$\tilde{\boldsymbol{g}}^{\alpha}_{i} \equiv \textbf{g}_{i}^{(k_{\alpha})}/\tilde{m}_{k_{\alpha}} - {\textbf{g}}_{i}^{(l_{\alpha})}/\tilde{m}_{l_{\alpha}}$, 
since by Eqs. (\ref{LF1}) and (\ref{LF2}); $\textbf{r}^{\alpha}_{i+1}$ 
= $\textbf{r}^{(k_{\alpha})}_{i+1} - \textbf{r}^{(l_{\alpha})}_{i+1}$
= $\textbf{r}_{i}^{(k_{\alpha})} - \textbf{r}_{i}^{(l_{\alpha})} + \boldsymbol{\delta}_{i+1/2}^{(k_{\alpha})} - \boldsymbol{\delta}_{i+1/2}^{(l_{\alpha})}$
= $\textbf{r}^{\alpha}_{i} + \boldsymbol{\delta}^{\alpha}_{i-1/2} + l_{0}\lambda_{i}\tilde{\textbf{f}}_{i}^{\alpha} 
  + l_{0}\tilde{\textbf{g}}_{i}^{\alpha}$, it follows that 

\begin{align}
  U_{i-1} - \textbf{F}_{i}\cdot\Big[2\boldsymbol{\Delta}_{i-1/2} + l_{0}\lambda_{i}\tilde{\textbf{F}}_{i} + l_{0}\tilde{\textbf{G}}_{i} \Big] - U_{0} & = 0,\label{taylor1} \\
  \Big[ \textbf{r}^{\alpha}_{i} + \boldsymbol{\delta}^{\alpha}_{i-1/2} + l_{0}\lambda_{i}\tilde{\textbf{f}}_{i}^{\alpha} 
  + l_{0}\tilde{\textbf{g}}_{i}^{\alpha} \Big]^{2} - C^{2}_{\alpha} & = 0,\,\ (\alpha = 1,...,G).\label{taylor2}
\end{align}
The above coupled quadratic equations for the Lagrangian multipliers are now solved following the produce of the \textit{MILC-SHAKE} algorithm\cite{milcshake}, which 
starts by neglecting the second order terms in the Lagrangian multipliers and solving the resulting linear equations. Afterwards, the second order 
terms are taken into account in an iterative manner - the details of which are described below. 

For each integration step $i$, the linearized equations are given by

\begin{equation}\label{linear}
  \textbf{A}_{i}\boldsymbol{\lambda}_{i} = \textbf{b}_{i},
\end{equation}
where $\textbf{A}_{i}$ is a $(G+1) \times (G+1)$ matrix, $\boldsymbol{\lambda}_{i} \equiv \{\lambda_{i}, \Lambda_{1 i}, ..., \Lambda_{G i} \}$, and 
$\textbf{b}_{i}$ a $G+1$ column vector. We start by calculating explicitly the first few elements of the matrix $\textbf{A}_{i}$.
$A_{11}$ consists merely of the factor infront of $\lambda_{i}$ in Eq. (\ref{taylor1}), i.e., $A_{11} = -l_{0}\tilde{\textbf{F}}_{i}\cdot\textbf{F}_{i}$. 
The second element $A_{12}$ appears after expansion of the dot product $\textbf{F}_{i} \cdot \tilde{\textbf{G}}_{i}$.
Noting that $\nabla_{\textbf{r}_{i}^{(k_{\alpha})}}\sigma_{\alpha} = 2\textbf{r}_{i}^{\alpha}$, we have $\textbf{F}_{i} \cdot \tilde{\textbf{G}}_{i}$ =
$\textbf{f}_{i}^{(1)}\cdot \textbf{g}_{i}^{(1)}/\tilde{m}_{1} + ... + \textbf{f}_{i}^{(N)}\cdot \textbf{g}_{i}^{(N)}/\tilde{m}_{N}$
= 2$\Lambda_{1i}(\tilde{\textbf{f}}^{1}_{i}\cdot\textbf{r}_{i}^{1}) + ... + 2\Lambda_{Gi}(\tilde{\textbf{f}}^{G}_{i}\cdot\textbf{r}_{i}^{G})$. The last equation follows as the Lagrangian multipliers 
appear in pairs, differing only by the sign from $\nabla_{\textbf{r}_{i}^{(k_{\alpha})}}\sigma_{\alpha}$ and the term $\textbf{f}_{i}^{(k_{\alpha})}/\tilde{m}_{k_{\alpha}}$. We thus 
find $A_{12}$ = $-2l_{0}\tilde{\textbf{f}}^{1}_{i}\cdot\textbf{r}_{i}^{1}$, $A_{13} = -2l_{0}\tilde{\textbf{f}}^{2}_{i}\cdot\textbf{r}_{i}^{2}$, etc. 
In the second row of $\textbf{A}_{i}$, the short-hand notation $\textbf{s}^{\alpha}_{i} \equiv \textbf{r}_{i}^{\alpha} + \boldsymbol{\delta}^{\alpha}_{i-1/2}$ is introduced, making
 $A_{21}$ = $2l_{0}(\textbf{s}^{1}_{i} \cdot \tilde{\textbf{f}}^{1}_{i})$, i.e., the factor infront of $\lambda_{i}$ after squaring of the parentheses. The next element $A_{22}$ 
appears after expanding $\textbf{s}_{i}^{1} \cdot \tilde{\textbf{g}}^{1}_{i}$ 
= $\boldsymbol{s}^{1}_{i}\cdot\sum_{\beta}\Lambda_{\beta i}(\frac{1}{\tilde{m}_{k_{1}}}\nabla_{\textbf{r}_{i}^{(k_{1})}}\sigma_{\beta} - \frac{1}{\tilde{m}_{l_{1}}}\nabla_{\textbf{r}_{i}^{(l_{1})}}\sigma_{\beta})$. 
In this sum, we identify the factor in front of $\Lambda_{1i}$, giving $A_{22}$ = $2l_{0}\boldsymbol{s}^{1}_{i}\cdot(\frac{1}{\tilde{m}_{k_{1}}}\nabla_{\textbf{r}_{i}^{(k_{1})}}\sigma_{1} 
- \frac{1}{\tilde{m}_{l_{1}}}\nabla_{\textbf{r}_{i}^{(l_{1})}}\sigma_{1})$, and similarly for the remaining elements of the second row.

Altogether, the elements of $\textbf{A}_{i}$ are thus given by
\begin{align}
  \textbf{A}_{i} & = 2l_{0}
  \begin{pmatrix}
    -\tilde{\textbf{F}}_{i}\cdot\textbf{F}_{i} / 2 & -\tilde{\textbf{f}}_{i}^{1}\cdot \textbf{r}^{1}_{i} & \cdots &  -\tilde{\textbf{f}}_{i}^{G}\cdot \textbf{r}^{G}_{i} \\ 
    \boldsymbol{s}^{1}_{i} \cdot \tilde{\textbf{f}}_{i}^{1} & \boldsymbol{s}^{1}_{i}\cdot(\frac{1}{\tilde{m}_{k_{1}}}\nabla_{\textbf{r}_{i}^{(k_{1})}}\sigma_{1} 
    - \frac{1}{\tilde{m}_{l_{1}}}\nabla_{\textbf{r}_{i}^{(l_{1})}}\sigma_{1}) & \cdots & 
    \boldsymbol{s}^{1}_{i}\cdot(\frac{1}{\tilde{m}_{k_{1}}}\nabla_{\textbf{r}_{i}^{(k_{1})}}\sigma_{G} 
    - \frac{1}{\tilde{m}_{l_{1}}}\nabla_{\textbf{r}_{i}^{(l_{1})}}\sigma_{G})  \\ 
    \vdots  & \vdots  & \vdots & \vdots \\
    \boldsymbol{s}^{G}_{i} \cdot \tilde{\textbf{f}}_{i}^{G} & \boldsymbol{s}^{G}_{i}\cdot(\frac{1}{\tilde{m}_{k_{G}}}\nabla_{\textbf{r}_{i}^{(k_{G})}}\sigma_{1} 
    - \frac{1}{\tilde{m}_{l_{G}}}\nabla_{\textbf{r}_{i}^{(l_{G})}}\sigma_{1}) & \cdots & 
    \boldsymbol{s}^{G}_{i}\cdot(\frac{1}{\tilde{m}_{k_{G}}}\nabla_{\textbf{r}_{i}^{(k_{G})}}\sigma_{G} 
    - \frac{1}{\tilde{m}_{l_{G}}}\nabla_{\textbf{r}_{i}^{(l_{G})}}\sigma_{G})  
  \end{pmatrix}. 
\end{align}
The column vector $\textbf{b}_{i}$ consists of all  zeroth-order terms in Eqs. (\ref{taylor1}) and (\ref{taylor2}) 

\begin{align}
  \textbf{b}_{i} & = 
  \begin{pmatrix}
    U_{0} - U_{i-1} + 2\textbf{F}_{i}\cdot\boldsymbol{\Delta}_{i-1/2} \\
    C^{2}_{1} - (\textbf{s}^{1}_{i})^{2} \\
    \vdots \\ 
    C^{2}_{G} - (\textbf{s}^{G}_{i})^{2}
  \end{pmatrix}.
\end{align}
Turning now to the iteration procedure, the second-order terms in the Lagrangian multipliers (Eq. (\ref{taylor2})) are taken into account by iterating
the right-hand side of Eq. (\ref{linear}) via the scheme ($\alpha = 1, ..., G$)

\begin{equation}\label{ite}
  b_{\alpha}^{j+1} = b_{\alpha}^{j} + \Big[C_{\alpha}^{2} - \big((\textbf{r}_{i+1}^{\alpha})^{2} \big)^{j}\Big].
\end{equation}
The superscript $j$ refers here to iteration $j$, and $\big((\textbf{r}_{i+1}^{\alpha})^{2}\big)^{j}$ are the positions associated with
iteration $j$. The element $b_{0}$ is not updated as it derives from the 
constraint of constant potential energy. For each iteration $j$ the term $C_{\alpha}^{2} - \big((\textbf{r}_{i+1}^{\alpha})^{2}\big)^{j}$ is expected to become smaller as the bonds 
are satisfied better and better, and convergence was achieved within a few iterations\cite{milcshake}.

For each integration step $i$, the algorithm for determining the \textit{NVU} Lagrangian multipliers thus proceeds as follows 

\begin{enumerate}
\item The Lagrangian multipliers of iteration $j$, $(\boldsymbol{\lambda}_{i})^{j}$, are calculated from Eq. (\ref{linear}). 
\item $\big((\textbf{r}_{i+1}^{\alpha})^{2} \big)^{j}$ is calculated via Eqs. (\ref{LF1}) and (\ref{LF2}) using $(\boldsymbol{\lambda}_{i})^{j}$.
\item $\textbf{b}_{i}$ is updated via Eq. (\ref{ite}) from $\big((\textbf{r}_{i+1}^{\alpha})^{2} \big)^{j}$.
\item The above steps are repeated (starting iteration $j+1$) until convergence is established (we used a preset number of iterations, typically 3-5).
\end{enumerate}
How is constant step length $l_{0}$ ensured numerically? Generalizing the approach of Paper I we introduce a normalizing factor such that 

\begin{align}
  \boldsymbol{\delta}_{i+1/2}^{(k)} & = l_{0}\frac{\boldsymbol{\chi}^{(k)}_{i-1/2}}{\sqrt{\sum_{k}\tilde{m}_{k}(\boldsymbol{\chi}_{i-1/2}^{(k)})^{2}}}, \\
  \textbf{r}_{i+1}^{(k)} & = \textbf{r}_{i}^{(k)} + \boldsymbol{\delta}_{i+1/2}^{(k)},
\end{align}
where 

\begin{equation}
  \boldsymbol{\chi}_{i-1/2}^{(k)} \equiv \boldsymbol{\delta}_{i-1/2}^{(k)}  + \frac{l_{0}}{\tilde{m}_{k}}\lambda_{i}\textbf{f}_{i}^{(k)} 
  +  \frac{l_{0}}{\tilde{m}_{k}}\textbf{g}_{i}^{(k)}.
\end{equation}
The normalizing factor is close to unity\cite{nvu1} and ensures trivially 
$\sum_{k}\tilde{m}_{k}(\boldsymbol{\delta}_{i+1/2}^{(k)})^{2}$ = $l^{2}_{0}$, i.e., that the step length is conserved. The algorithm is now absolutely stable, conserving 
potential energy, bond lengths, and step length for indefinitely long runs. The stability of the \textit{NVU} algorithm is tested numerically in Sec. \ref{testing}.

\subsection{Alternative determination of the \textit{NVU} Lagrangian multipliers}\label{alternative}

The previous section followed the traditional way of calculating the Lagrangian multipliers.
The \textit{NVU} Lagrangian multipliers may also be calculated by Taylor expanding the constraints $\sigma_{\alpha}$ in analogy to the method sketched above for the 
potential energy. In this way, the constraints of constant potential energy and constant bond lengths are treated on equal footing.
The set of equations to be solved is the following (recall that
$\textbf{R}_{i+1} - \textbf{R}_{i-1}$ = $2\boldsymbol{\Delta}_{i-1/2} + l_{0}\lambda_{i}\tilde{\textbf{F}}_{i} + l_{0}\tilde{\textbf{G}}_{i}$)

\begin{align}
  U_{i-1} - \textbf{F}_{i} \cdot (\textbf{R}_{i+1} - \textbf{R}_{i-1}) - U_{0}  & = 0, \label{linear1} \\ 
  \sigma_{\alpha(i-1)} + \nabla_{\textbf{R}_{i}}\sigma_{\alpha i} \cdot (\textbf{R}_{i+1} - \textbf{R}_{i-1}) - C^{2}_{\alpha}  & = 0,\,\ (\alpha = 1,...,G). \label{linear2}
\end{align}
The determination of the Lagrangian multipliers is linear and thus no iterations are needed. The bond 
constraints $\sigma_{\alpha}$ are obeyed to the same order $O(l_{0}^{3})$ as the constraint of constant potential energy. The sampling properties of this novel, alternative 
determination method is tested briefly in Sec. \ref{sampling}. It appears to be a promising new way of determining the Lagrangian multipliers in connection with rigid bonds, 
which might also be useful for standard bond-constraint \textit{NVE} or \textit{NVT} simulations.

\section{Simulation details and model systems}\label{rigidmodel}

We investigated three systems: The asymmetric dumbbell model, the Lewis-Wahnstr{\"o}m OTP model, and rigid SPC/E water. For all simulated pair potentials the shifted-force 
truncation scheme was applied at a cut-off radius $r_{c}$. If the pair potential is $v(r)$ and the pair force is $f(r)=-v'(r)$, the shifted force is given by\cite{tildesley,FCS1} 

\begin{equation}
  f_{\rm SF}(r)\,=\,
  \begin{cases}  f(r)-f(r_c) & \text{if}\,\, r<r_c\,, \\ 
    0 &\text{if}\,\, r>r_c\,.
  \end{cases} 
\end{equation}
This corresponds to using the following pair potential below $r_c$: $v_{\rm SF}(r) = v(r) - v'(r_c) ( r-r_c) - v(r_c)$. All simulations were 
performed with the \textit{NVT} and \textit{NVU} algorithms. Recall that \textit{NVE} and \textit{NVT} dynamics give equivalent results\cite{evans1985}; for this reason no 
simulations are presented for \textit{NVE} dynamics.
The RUMD code\cite{rumd} was used for molecular dynamics simulations (an optimized open-source GPU code). The \textit{NVT} ensemble is 
generated via the Nosé-Hoover algorithm\cite{nose,hoover,nvttoxvaerd}, and the bonds held fixed using the time-reversible constraint 
algorithm of Refs. \onlinecite{toxconstraintnve,toxconstraintnph}. The \textit{NVU} algorithm is described in Sec. \ref{rigidalgo}. The starting files for \textit{NVU} dynamics 
were taken from an equilibrated \textit{NVT} simulation. The positions and velocities of the \textit{NVT} configuration do not correspond perfectly 
to motion on $\omega$, since the potential energy and step length are not those of $U_{0}$ and $l_{0}$, respectively. As all three constraints are to be satisfied 
simultaneously, this results in numerical problems when starting the simulation from the particular \textit{NVT} configuration.
A more gentle procedure is thus applied, where the atomic \textit{NVU} algorithm is used for a couple of integration steps to ensure the values of $U_{0}$ and $l_{0}$. Afterwards, the 
rigid-bond \textit{NVU} algorithm is used.

\subsection{\textit{NVU} iteration procedure}

The quadratic equations (Eq. (\ref{ite})) were iterated with a fixed number of iterations (between 3 and 5).
The linear systems were solved utilizing CUSP\cite{cusp}, a library for solving systems of linear equations on the \textit{GPU}. More specifically, 
the stabilized biconjugate gradient algorithm with a Jacobi preconditioner\cite{recipes} was used with the initial value $\boldsymbol{\lambda}_{i} = 0$.
The relative tolerance $\tau$ for the asymmetric dumbbell and Lewis-Wahnstr{\"o}m OTP models was chosen as $\tau = 10^{-7}$ and for rigid SPC/E water as $\tau = 3 \cdot 10^{-7}$. 
A larger tolerance was chosen for rigid SPC/E water due to convergence issues in connection with shifted-force Coulomb interactions (see below).

The maximum number of allowed iterations was 50. A restart scheme was applied when the solver did not converge within the chosen tolerance. In this case the solver (and quadratic iteration) was 
restarted from the partially estimated ''solution'' adding $2 \cdot 10^{-7}$ to the tolerance. It should be noted that the stabilized biconjugate 
gradient algorithm may get ''trapped'', resulting in a break-down of the CUSP linear solver. If this happens, it is 
detected by our program, and the solver and quadratic iteration are restarted, with a 
smaller number (10) of maximum allowed iterations for the solver. 

\subsection{The asymmetric dumbbell}

The asymmetric dumbbell model\cite{moleculeshidden} consists of a large ($A$) and a small ($B$) Lennard-Jones (LJ) particle, rigidly bonded  with bond distance of $r_{AB} = 0.29/0.4963$ (here and 
henceforth units are given in LJ units referring to the $A$ particle such that $\sigma_{AA}$ = 1, $\epsilon_{AA}$ = 1, and $m_{A}$ = 1). The asymmetric dumbbell 
model has $\sigma_{BB}=0.3910/0.4963$, $\epsilon_{BB}=0.66944/5.726$, and $m_{B}=15.035/77.106$. The $AB$ interaction between different 
molecules is determined by the Lorentz-Berthelot mixing rule\cite{tildesley}.
$n$ = 500 molecules (here and henceforth $n$ denotes the number of molecules and $N$ the number of atoms) were used in the simulations 
with a pair-potential cut-off of $r_{c} = 2.5$. The step length $l_{0}$ was fixed in the range 0.125-0.138 depending on the state point.

Simulations were also performed where the rigid bonds were replaced with stiff harmonic springs. The spring constant was $k = 3000$, while all other model parameters remained unchanged.

\subsection{Lewis-Wahnstr{\"o}m OTP}

The Lewis-Wahnstr{\"o}m OTP model\cite{otp2} consists of three identical LJ particles rigidly bonded in an isosceles triangle with sides of $r_{AA} = 1$ and 
top angle of $75^\circ$. All parameters (including the masses) are unity for the OTP model. $n$ = 320 molecules were simulated and a pair-potential cut-off of $r_{c} = 2.5$ was used.
The step length was 0.100.

\subsection{SPC/E Water}

The SPC/E water model\cite{spce} is an isosceles triangle with sides $r_{OH} = 1/3.166$ and top angle $109.47^{\circ}$. The $OO$ intermolecular interactions are given by the 
LJ pair potential ($\epsilon_{OO}=1$, $\sigma_{OO}=1$, and $m_{O}=15.9994/1.00794$). The three particles are 
charged with $q_{O}=-22.0$ and $q_{H}=|q_{O}|/2$. $n$ = 2000 molecules were simulated and a pair-potential cut-off 
of $r_{c} = 6.28$ for both LJ and Coulomb interactions was applied\cite{shiftedforceCoulomb,shiftedforceHansen}. The step length was fixed in the range 0.06-0.07 
depending on the state point. For this system the numerical 
stability is surprisingly sensitive to the cut-off used in the Coulomb interactions, but a larger shifted-force cut-off improves this behavior\cite{shiftedforceHansen}.

\section{Testing the stability of the rigid-bond $NVU$ algorithm}\label{testing}

This section tests the conservation properties of the rigid-bond \textit{NVU} algorithm. Table \ref{ustepconservation} shows the potential energy, the deviation of bond lengths, and step length
as functions of integration step number for Lewis-Wahnstr{\"o}m OTP at $\rho$ = 0.329 and T = 0.700. It is clear that these quantities are conserved by the algorithm 
and that no drift occurs. The step length is conserved to the highest accuracy since it is not prone to numerical error in determining the Lagrangian multipliers.

\begin{figure}[H]
  \centering
  \includegraphics[width=70mm]{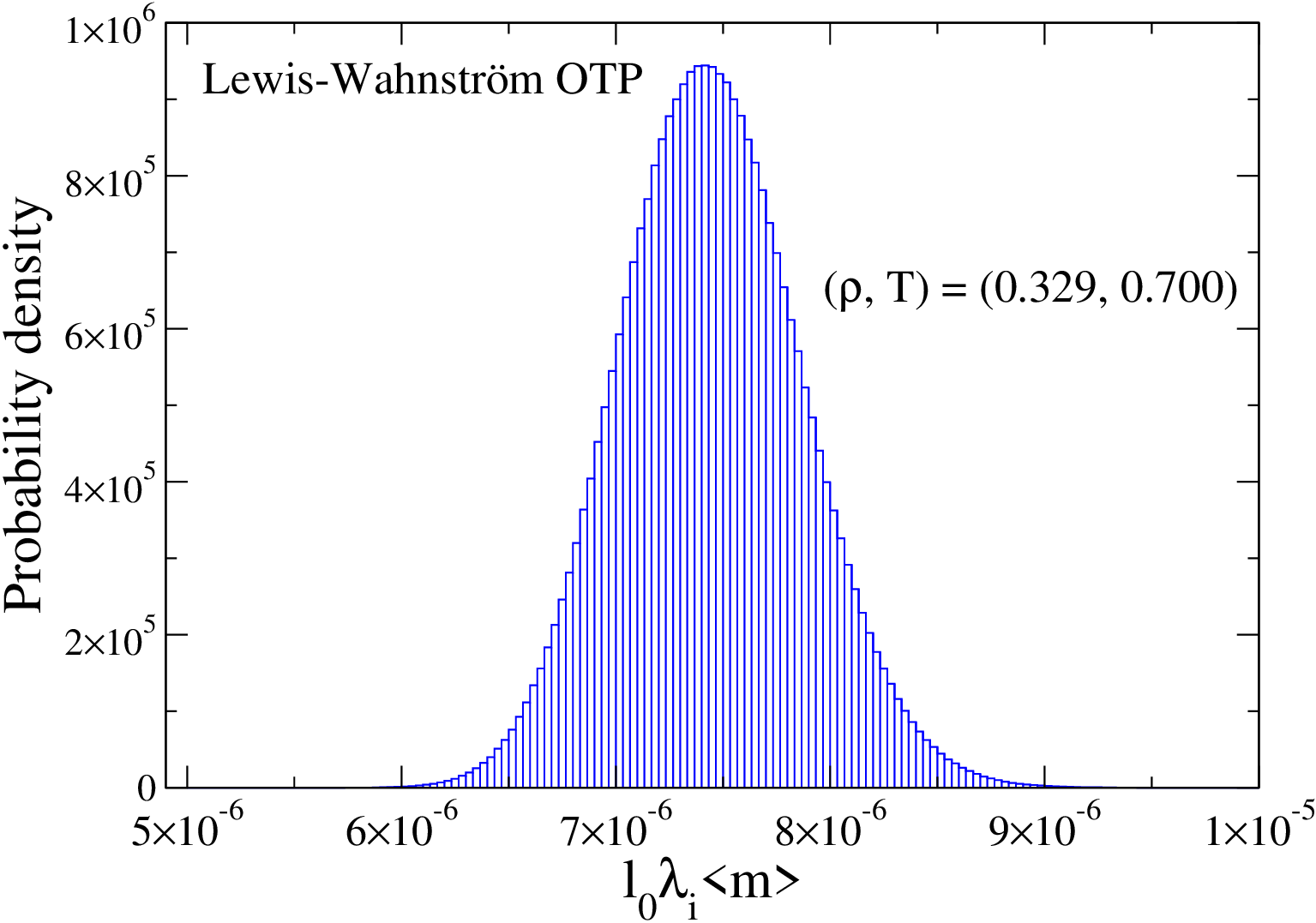}
  \caption{The probability density of the ''time step'' length $(\Delta t_{i,NVU})^{2} \equiv l_{0}\lambda_{i}\langle m \rangle$  of the rigid-bond \textit{NVU} algorithm for 
    Lewis-Wahnstr{\"o}m OTP at $\rho$ = 0.329 and T = 0.700. $n = 320$ molecules were used in the simulations.}
  \label{lambdadist}
\end{figure}

\begin{table}[H]
  \centering
  \begin{tabular}
    {|| p{4.0cm} || p{4.0cm} | p{4.0cm} | p{4.0cm} || }
    \hline
    \hline
    Integration steps & $U$ / $N$ & $(1/G\sum_{\alpha}(r^{\alpha}$ $-$ $C_{\alpha})^{2})^{1/2}$ & $\sum_{k}\tilde{m}_{k}(\boldsymbol{\delta}_{i+1/2}^{(k)})^{2}$ \\
    \hline
    \hline
    $10^{1}$  & -4.42550 & 2.81207 $\cdot 10^{-7}$ & 0.0999999 \\
    $10^{2}$  & -4.42552 & 3.03535 $\cdot 10^{-7}$ & 0.1000000 \\
    $10^{3}$  & -4.42552 & 2.81128 $\cdot 10^{-7}$ & 0.1000000 \\
    $10^{4}$  & -4.42552 & 2.95078 $\cdot 10^{-7}$ & 0.1000000 \\
    $10^{5}$  & -4.42550 & 3.08793 $\cdot 10^{-7}$ & 0.1000000 \\
    $10^{6}$  & -4.42551 & 2.90477 $\cdot 10^{-7}$ & 0.1000000 \\
    \hline
    \hline 
  \end{tabular}
  \caption{Potential energy, deviation of bond lengths and step length as functions of integration step number in the \textit{NVU} algorithm for Lewis-Wahnstr{\"o}m OTP ($\rho$ = 0.329, T = 0.700). 
    Single-precision floating-point arithmetic was used for the simulations.}
  \label{ustepconservation}
\end{table}
Figure \ref{lambdadist} shows the distribution of the term $l_{0}\lambda_{i}\langle m \rangle$ in Eq. (\ref{some}) (recall $\tilde{m}_{k} = m_{k}/\langle m \rangle$). 
In \textit{NVU} dynamics there is, as such, no notation of time;
a geodesic on the manifold can be traversed with any velocity. Comparing the \textit{NVU} algorithm of Eq. (\ref{some}) 
to the rigid-bond Verlet algorithm\cite{shake} $\br^{(k)}_{i+1} \,=\, 2\br^{(k)}_i - \br^{(k)}_{i-1} + ((\Delta t)^{2}/m_{k})[\textbf{f}^{(k)}_i + \textbf{g}^{(k)}_i]$, 
we can define the term $l_{0}\lambda_{i}\langle m \rangle$ as 
a varying ''time step'' length of the \textit{NVU} algorithm (see also Paper II), i.e.,

\begin{equation}\label{time}
  (\Delta t_{i,NVU})^{2} \equiv l_{0}\lambda_{i}\langle m \rangle. 
\end{equation}
The integration steps of the \textit{NVU} algorithm are thus henceforth referred to as ''time steps''. The average of 
Eq. (\ref{time}) is used in Sec. \ref{sampling} when comparing to \textit{NVT} dynamics. As was the case for the atomic \textit{NVU} 
algorithm (Paper I), $l_{0}\lambda_{i}\langle m \rangle$ is Gaussian distributed for large systems and its relative variation decreases as 
the number of particles increases. It thus becomes a better and better approximation to treat this term as constant, implying equivalent 
sampling properties  of \textit{NVU} and \textit{NVE} dynamics also when rigid bonds are included in the simulations.\newline \newline

\section{Sampling properties of the rigid-bond $NVU$ algorithm}\label{sampling}

The \textit{NVU} algorithm is now compared to \textit{NVT} dynamics for the three different models.
First, we consider the asymmetric dumbbell model\cite{moleculeshidden}, both rigid and flexible. Afterwards, the {Lewis-Wahnstr{\"o}m OTP model\cite{otp2}, and 
finally rigid SPC/E water\cite{spce}.

\subsection{The asymmetric dumbbell model}

In Figs. \ref{dumbrigid}(a) and (b) are shown, respectively, the molecular center-of-mass (CM) radial distribution functions and the CM incoherent intermediate scattering functions for the rigid 
asymmetric dumbbell model\cite{moleculeshidden} for different temperatures at $\rho = 0.932$. The black circles and curves
give \textit{NVT} simulation results while the red crosses give the \textit{NVU} simulation results. The two radial distribution functions in Fig. \ref{dumbrigid}(a)
agree very well, and this is also the case for the dynamics in Fig. \ref{dumbrigid}(b). \newline \newline

\begin{figure}[H]
  \begin{minipage}[t]{0.45\linewidth}   
    \centering
    \includegraphics[width=70mm]{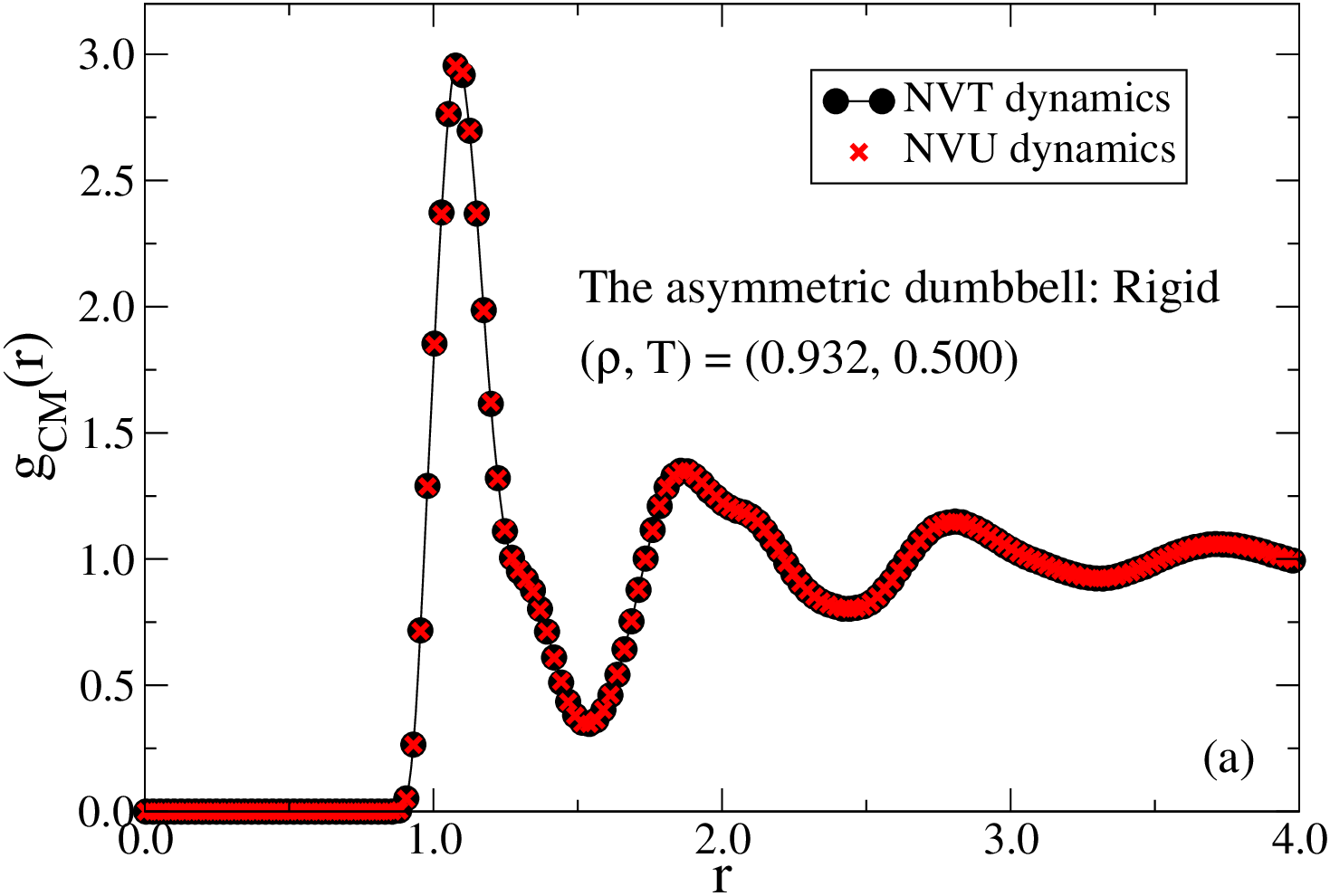}
  \end{minipage}
  \hspace{0.5cm}
  \begin{minipage}[t]{0.45\linewidth}
    \includegraphics[width=70mm]{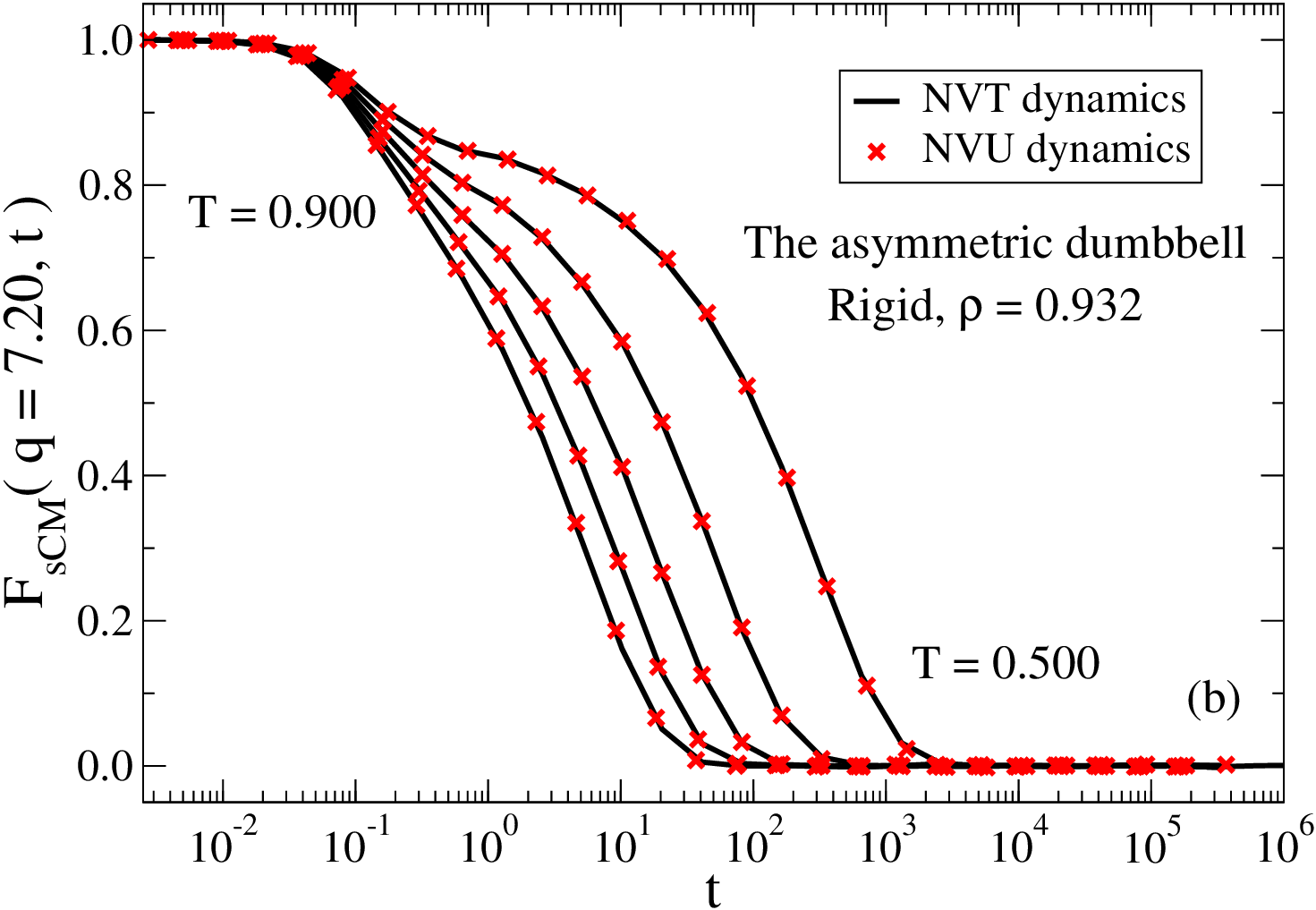}
    \centering
  \end{minipage}
  \caption{Comparison of structure and dynamics in \textit{NVU} and \textit{NVT} simulations of the rigid asymmetric dumbbell model. 
    The black circles and curves give \textit{NVT}, the red crosses \textit{NVU} simulation results. (a) The molecular CM radial 
    distribution functions at $\rho$ = 0.932 and T = 0.500. (b) The molecular CM incoherent intermediate scattering functions 
    at $\rho$ = 0.932 and T = 0.500, 0.600, 0.700, 0.800, 0.900. }
  \label{dumbrigid}
\end{figure}
For reference, we also simulated (Fig. \ref{dumbflex}) the corresponding quantities for the flexible-bond asymmetric dumbbell model at the state points of Fig. \ref{dumbrigid}. Again, there 
is a very good agreement between \textit{NVU} and \textit{NVT} dynamics. \newline \newline

\begin{figure}[H]
  \begin{minipage}[t]{0.45\linewidth}   
    \centering
    \includegraphics[width=70mm]{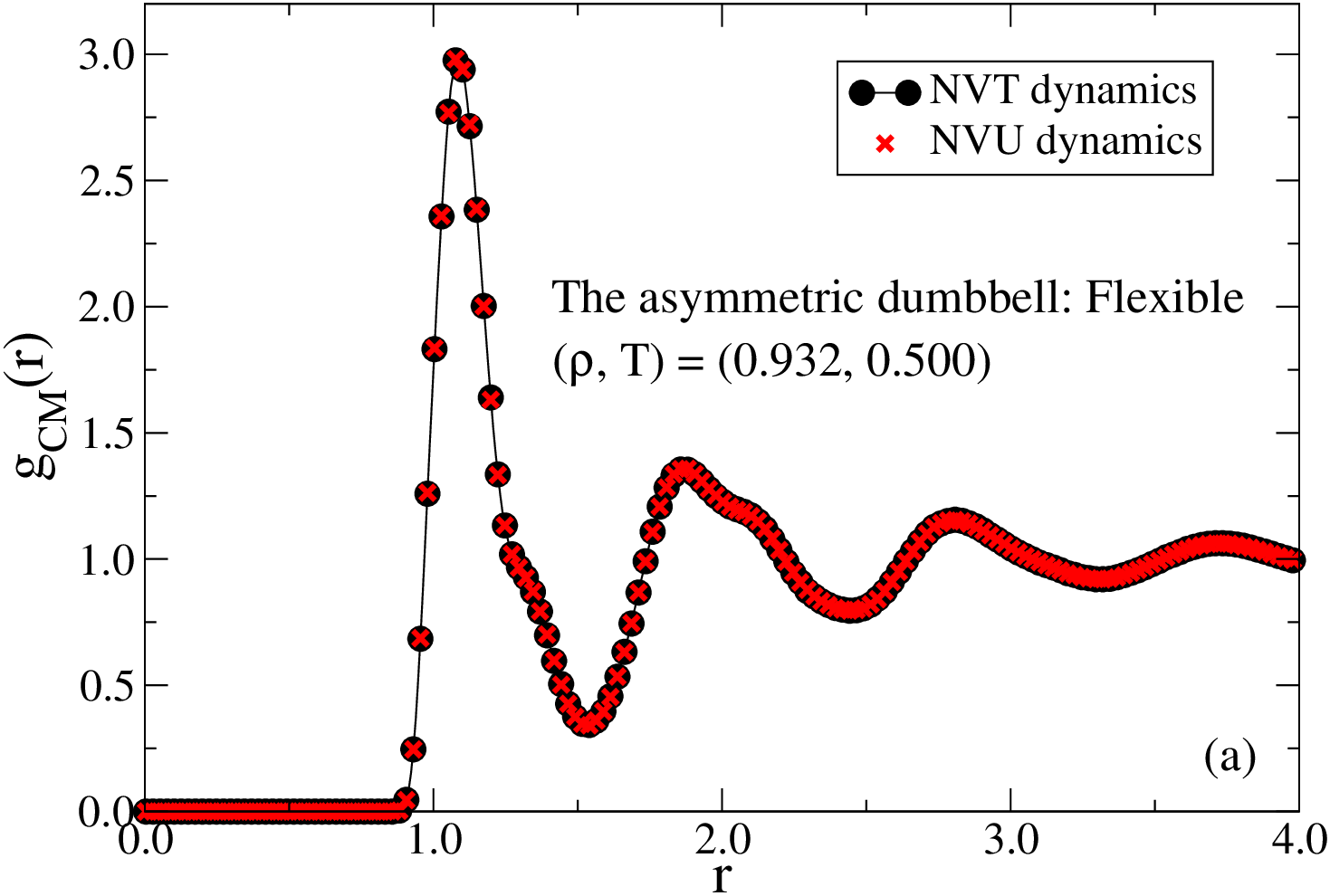}
  \end{minipage}
  \hspace{0.5cm}
  \begin{minipage}[t]{0.45\linewidth}
    \includegraphics[width=70mm]{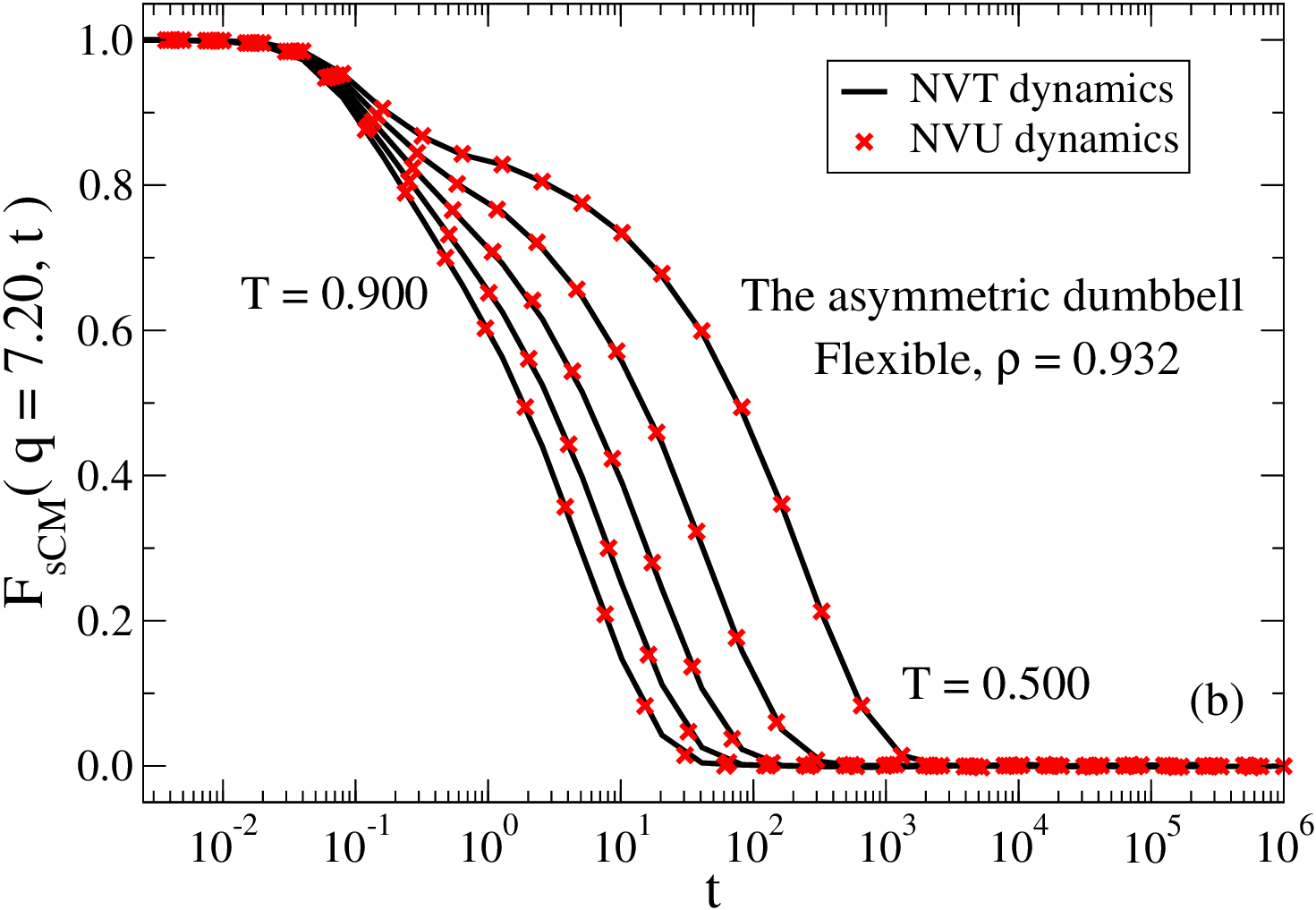}
    \centering
  \end{minipage}
  \caption{Comparison of structure and dynamics in \textit{NVU} and \textit{NVT} simulations of the flexible-bond asymmetric dumbbell model. 
    The black circles and curves give \textit{NVT}, the red crosses \textit{NVU} simulation results. The same state points as in Fig. \ref{dumbrigid} were simulated. 
    (a) The molecular CM radial distribution functions at $\rho$ = 0.932 and T = 0.500. (b) The molecular CM incoherent intermediate scattering functions 
    at $\rho$ = 0.932 and T = 0.500, 0.600, 0.700, 0.800, 0.900. }
  \label{dumbflex}
\end{figure}

\subsection{Lewis-Wahnstr{\"o}m OTP}

We show in Figs. \ref{otp}(a) and (b), respectively, the molecular CM radial distribution functions and CM incoherent intermediate scattering functions for 
the Lewis-Wahnstr{\"o}m OTP model\cite{otp2}. The same symbols and meanings as in the preceding section are used. Again, the \textit{NVU} and \textit{NVT} simulations agree very well for both 
structure and dynamics. \newline \newline

\begin{figure}[H]
  \begin{minipage}[t]{0.45\linewidth}   
    \centering
    \includegraphics[width=70mm]{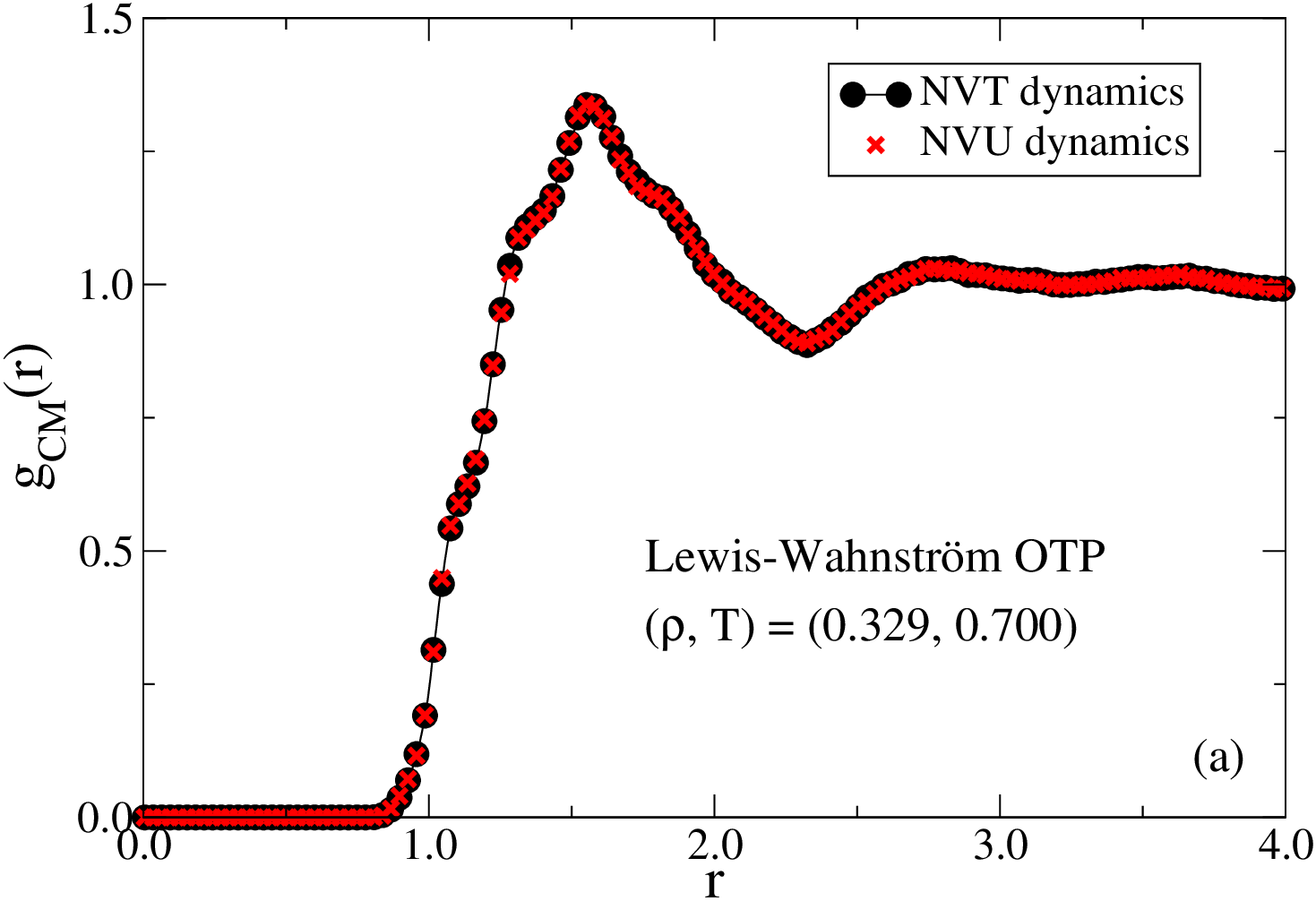}
  \end{minipage}
  \hspace{0.5cm}
  \begin{minipage}[t]{0.45\linewidth}
    \includegraphics[width=70mm]{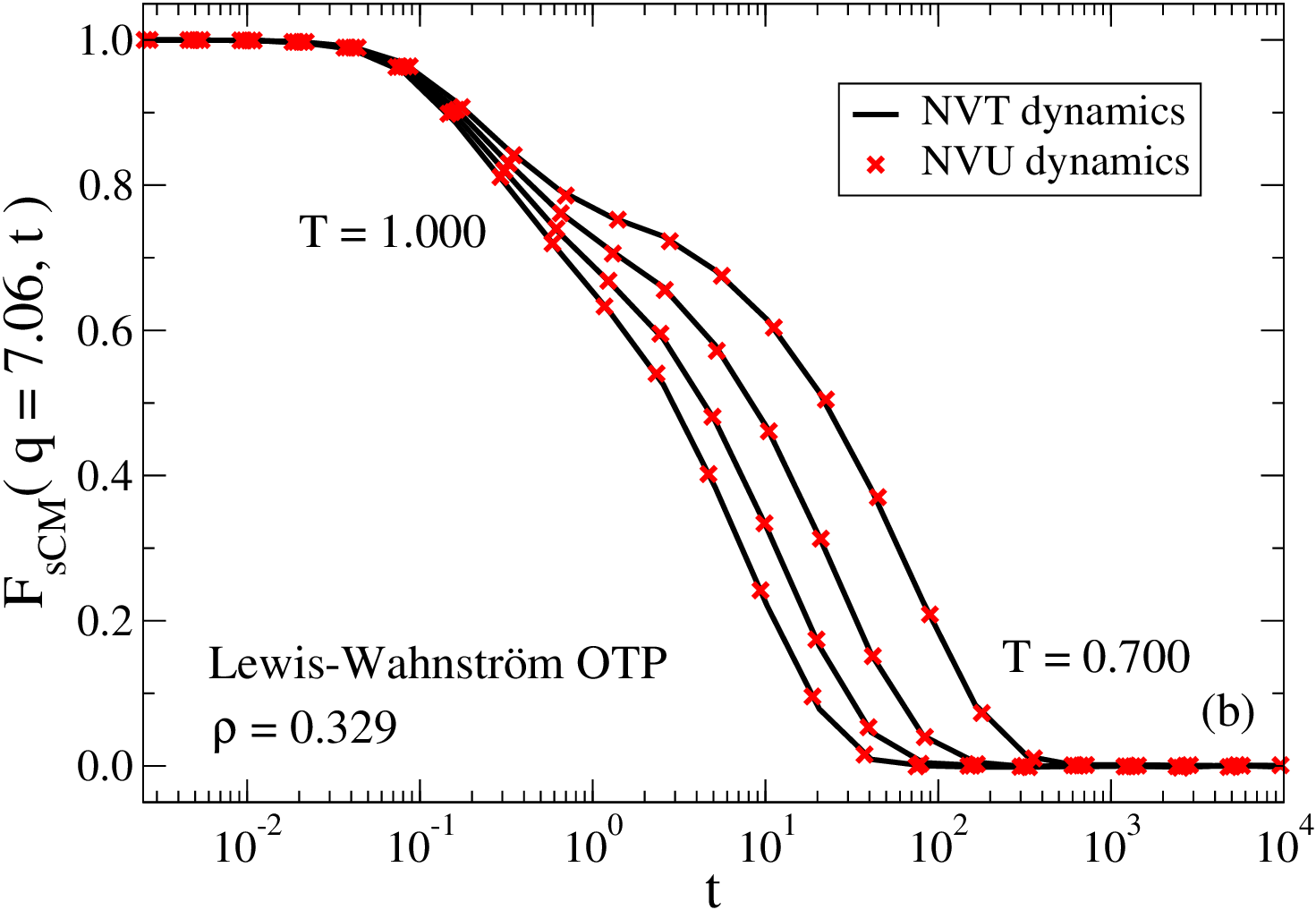}
    \centering
  \end{minipage}
  \caption{Comparison of center-of-mass structure and dynamics in \textit{NVU} and \textit{NVT} simulations of the Lewis-Wahnstr{\"o}m OTP model. 
    The black circles and curves give \textit{NVT}, the red crosses \textit{NVU} simulation results. (a) The molecular CM radial 
    distribution functions at $\rho$ = 0.329 and T = 0.700. (b) The molecular CM incoherent intermediate scattering functions 
    at $\rho$ = 0.329 and T = 0.700, 0.800, 0.900, 1.000. }
  \label{otp}
\end{figure}
For comparison, we also show in Fig. \ref{otp1} the corresponding particle quantities for the OTP model. \newline \newline

\begin{figure}[H]
  \begin{minipage}[t]{0.45\linewidth}   
    \centering
    \includegraphics[width=70mm]{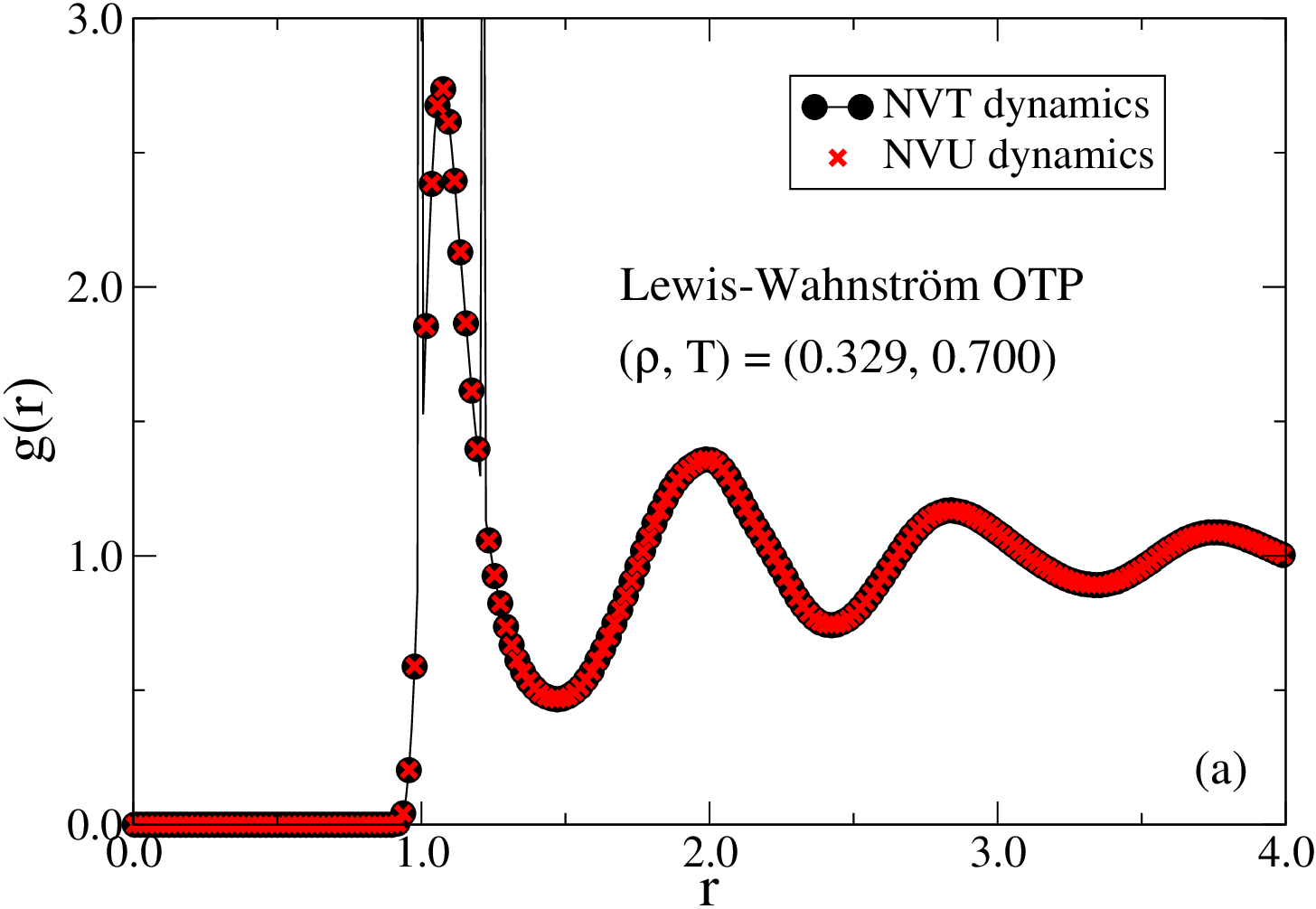}
  \end{minipage}
  \hspace{0.5cm}
  \begin{minipage}[t]{0.45\linewidth}
    \includegraphics[width=70mm]{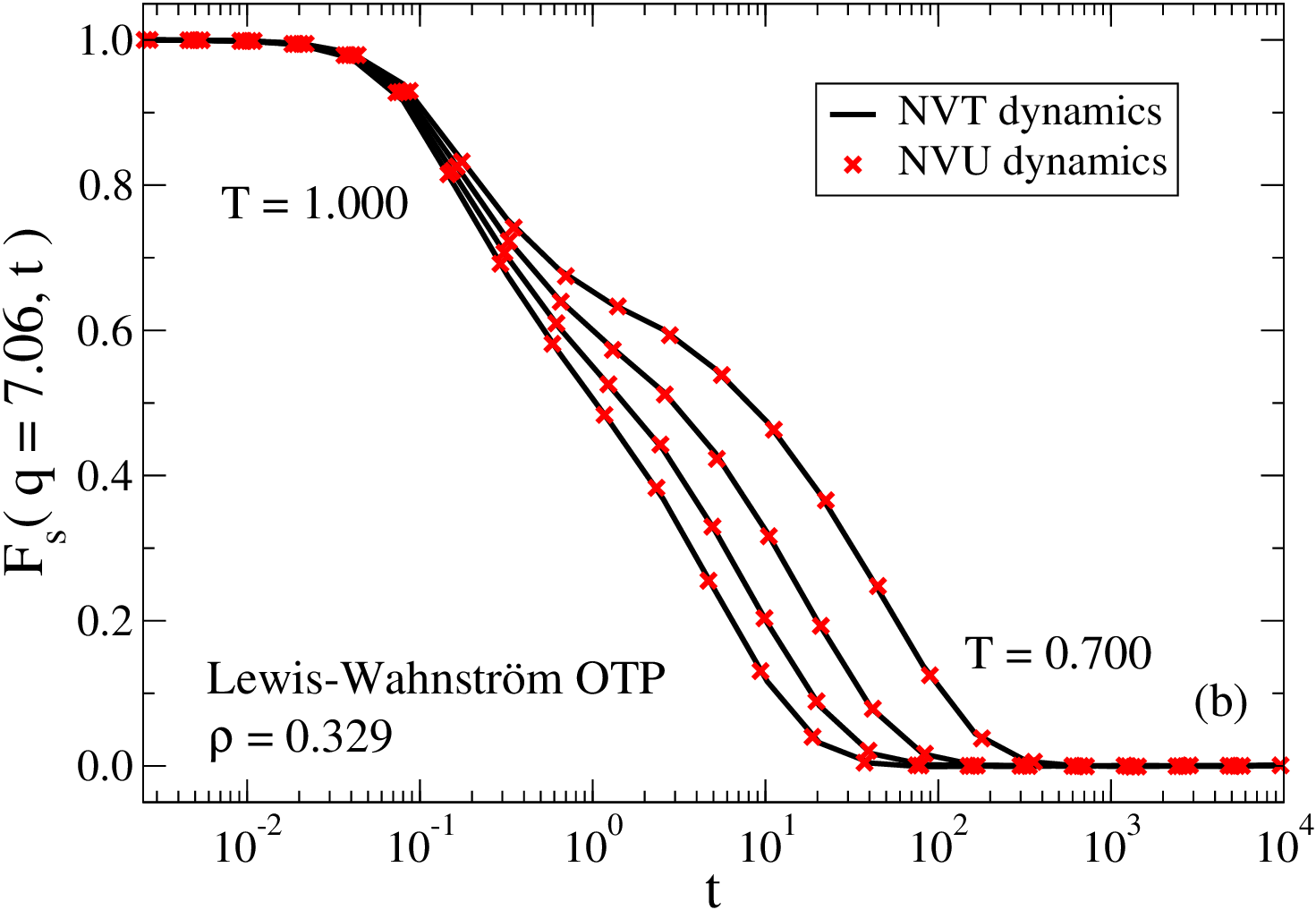}
    \centering
  \end{minipage}
  \caption{Comparison of particle structure and dynamics in \textit{NVU} and \textit{NVT} simulations for the Lewis-Wahnstr{\"o}m OTP model. 
    The black circles and curves give \textit{NVT}, the red crosses \textit{NVU} simulation results. (a) The particle radial 
    distribution function at $\rho$ = 0.329 and T = 0.700. (b) The particle incoherent intermediate scattering functions 
    at $\rho$ = 0.329 and T = 0.700, 0.800, 0.900, 1.000. }
  \label{otp1}
\end{figure}

\subsection{SPC/E Water}

Finally, we consider in Fig. \ref{spce} the same quantities as above for the rigid SPC/E water model\cite{spce}. Again, 
full equivalence between \textit{NVU} and \textit{NVT} dynamics is found. \newline \newline

\begin{figure}[H]
  \begin{minipage}[t]{0.45\linewidth}   
    \centering
    \includegraphics[width=70mm]{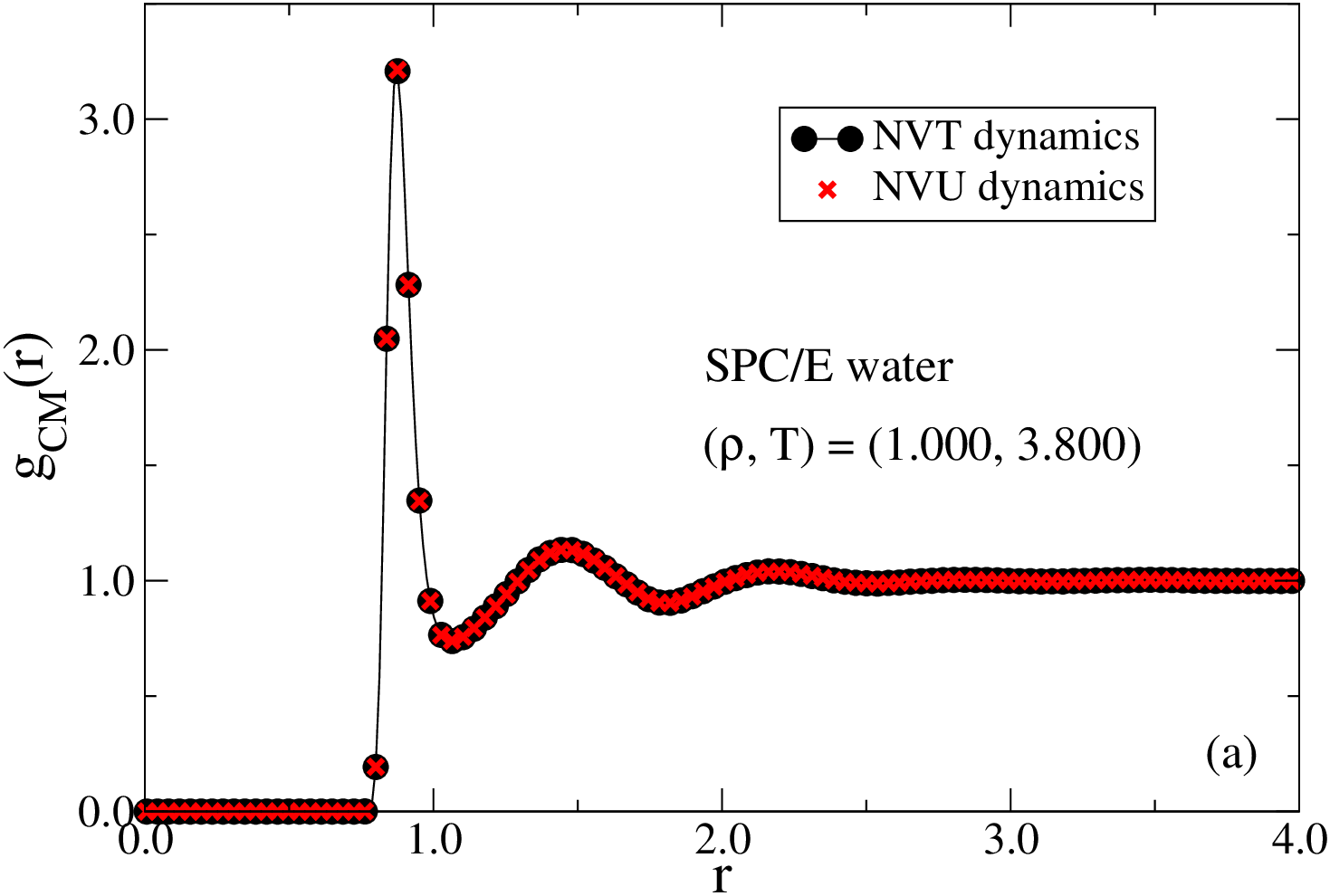}
  \end{minipage}
  \hspace{0.5cm}
  \begin{minipage}[t]{0.45\linewidth}
    \includegraphics[width=70mm]{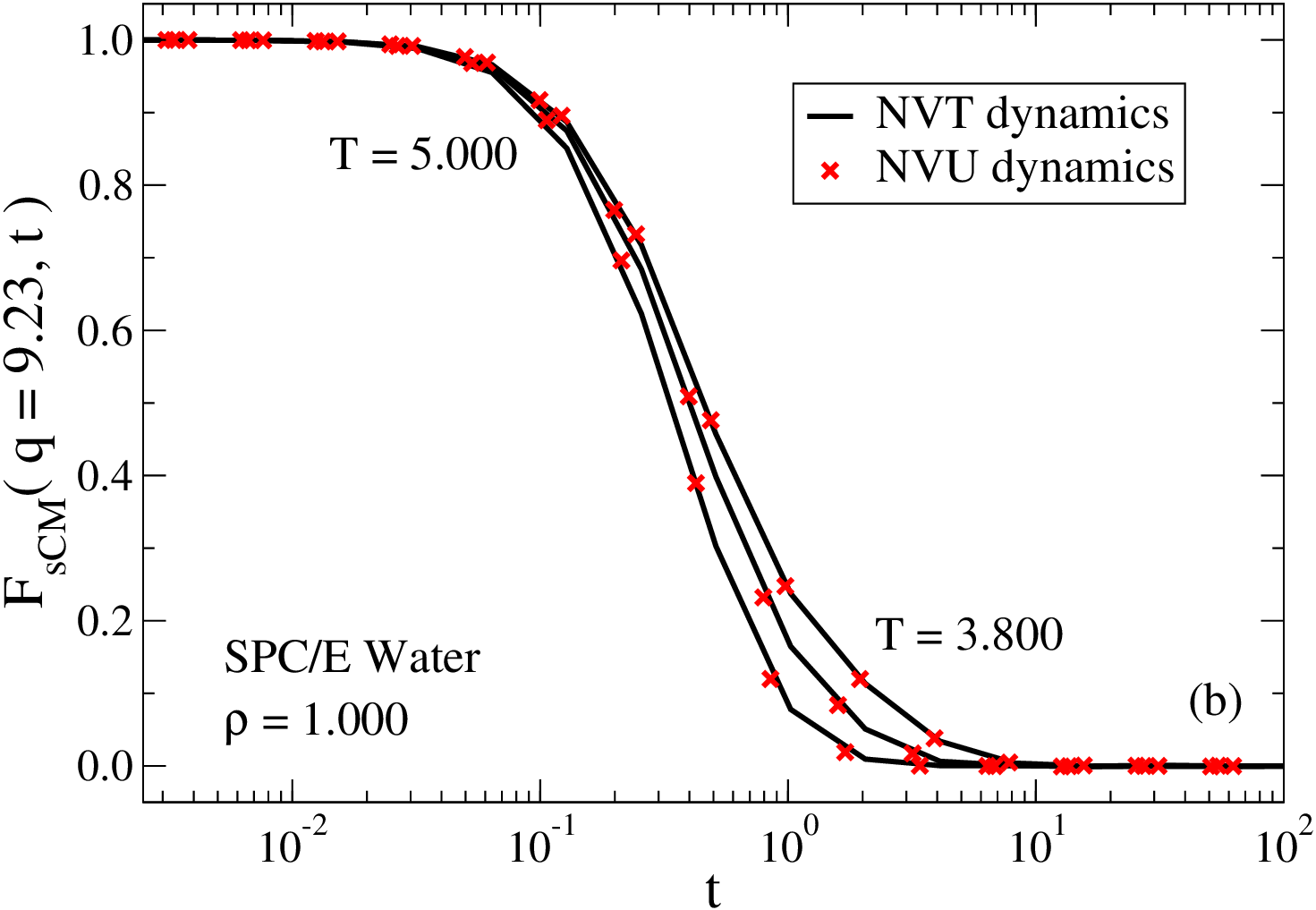}
    \centering
  \end{minipage}
  \caption{Comparison of structure and dynamics in \textit{NVU} and \textit{NVT} simulations of rigid SPC/E water. 
    The black circles and curves give \textit{NVT}, the red crosses \textit{NVU} simulation results. (a) The molecular CM radial 
    distribution functions at $\rho$ = 1.000 and T = 3.800. (b) The molecular CM incoherent intermediate scattering functions 
    at $\rho$ = 1.000 and T = 3.800, 4.200, 5.000. }
  \label{spce}
\end{figure}
The linear algorithm for determining the Lagrangian multipliers presented in Sec. \ref{alternative} (Eqs. (\ref{linear1}) and (\ref{linear2})) is 
tested in Fig. \ref{spcelinear} by probing the molecular CM radial distribution functions. \textit{NVU} and \textit{NVT} dynamics 
also here give identical results. \newline \newline

\begin{figure}[H]
  \centering
  \includegraphics[width=70mm]{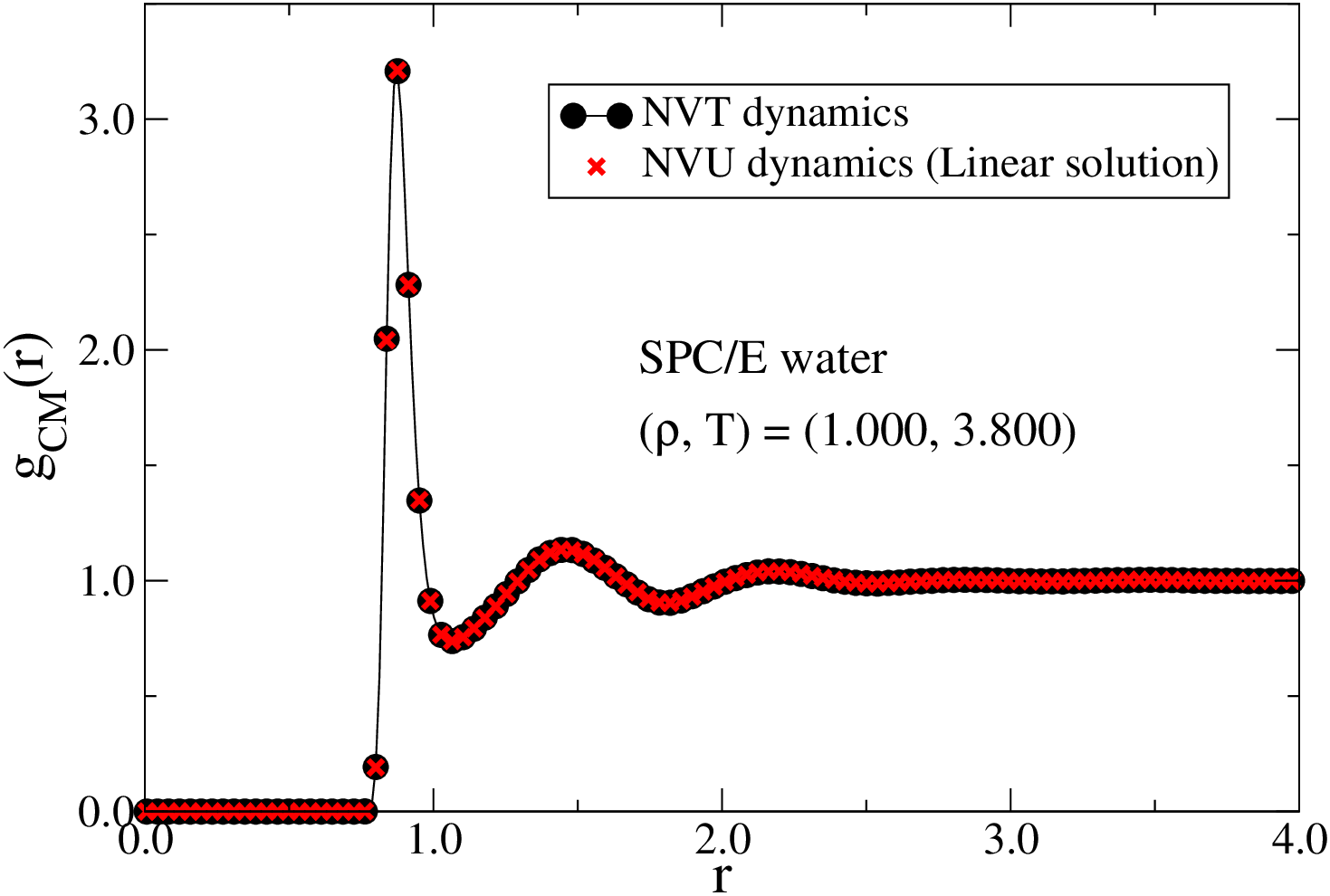}
  \caption{Comparison of structure in \textit{NVU} and \textit{NVT} simulations of rigid SPC/E water at $\rho$ = 1.000 and T = 3.800 applying the 
    linear method to determine the Lagrangian multipliers (Eqs. (\ref{linear1}) and (\ref{linear2})). The bond lengths are here 
conserved to order $10^{-6}$ in the standard deviation of the bonds (using single-precision).}
  \label{spcelinear}
\end{figure}
We conclude from the presented results that for sufficiently large molecular systems with 
flexible and/or rigid bonds, \textit{NVU} dynamics is equivalent to Nosé-Hoover \textit{NVT} dynamics (and, by implication, Newtonian \textit{NVE} dynamics).

\section{Summary}\label{sum}

\textit{NVU} dynamics is molecular dynamics at constant potential energy realized by tracing out a geodesic on the constant-potential-energy 
hypersurface $\Omega$ (Eq. (\ref{omega_def})). In Papers I and II\cite{nvu1,nvu2}, a ''basic'' and a ''stabilized'' atomic \textit{NVU} algorithm for simulating geodesics on $\Omega$ was developed. The 
basic \textit{NVU} algorithm has excellent stability; it is time-reversible and symplectic, and the stabilized algorithm was developed only to prevent accumulation of 
numerical error. It was found that atomic \textit{NVU} dynamics becomes equivalent to atomic \textit{NVE} dynamics in the thermodynamic limit. 

In this paper the  stabilized \textit{NVU} algorithm has been extended to simulate molecules at constant potential energy. Molecules are simulated by introducing rigid and/or flexible bonds in the models.
The atomic \textit{NVU} algorithm keeps the potential energy constant and can thus right away simulate flexible bonds. 
The focus here was on incorporating rigid bonds in the framework of \textit{NVU} dynamics, which leads to the introduction of additional Lagrangian multipliers beyond 
those of the constraint of constant potential energy. This is completely analogous to the approach for simulating rigid bonds in 
standard Newtonian \textit{NVE} dynamics\cite{shake,toxconstraintnve,toxconstraintnph}. In the \textit{NVU} algorithm, a set of coupled quadratic equations 
was constructed for calculating the Lagrangian multipliers and solved in an iterative manner as a linear system, a procedure developed for rigid-bond \textit{NVE} dynamics 
in the \textit{MILC-SHAKE} algorithm\cite{milcshake}. In addition, a set of linear equations was presented for calculating the Lagrangian multipliers, and appears to be a 
promising new way of simulating rigid bonds.

The rigid-bond \textit{NVU} algorithm reduces to the atomic 
\textit{NVU} algorithm when there are no rigid bonds. The algorithm was tested on three different model systems: the asymmetric dumbbell model, Lewis-Wahnstr{\"o}m OTP, and rigid SPC/E water.
The probed quantities in the simulation gave 
identical results to those of Nos$\acute{e}$-Hoover \textit{NVT} dynamics.
We conclude that also for molecular systems do \textit{NVU} dynamics become equivalent to \textit{NVE} dynamics in the thermodynamic limit (since \textit{NVE} and \textit{NVT} 
dynamics are known to give equivalent results\cite{evans1985}).

\acknowledgments 

The centre for viscous liquid dynamics ``Glass and Time'' is sponsored by the Danish National Research Foundation (DNRF). The authors are grateful to Ole J. Heilmann for pointing 
out the alternative method for determining the Lagrangian multipliers (Sec. \ref{alternative}).

\appendix

\section{Derivation of the atomic \textit{NVU} algorithm for the Hertzian metric}\label{hertz}

According to Newtonian dynamics, heavy particles move slower than light particles in thermal equilibrium. The standard Euclidean metric does not involve the particle 
masses, and thus applying this metric to geodesic motion for systems of varying masses will not produce dynamics equivalent to Newtonian dynamics 
in a thermal system. The mass-weighted metric of Hertz\cite{hertz1}, however, 
ensures that \textit{NVU} dynamics becomes equivalent to \textit{NVE} dynamics in the thermodynamic limit, as is clear from the 
derivation below. This metric is given by (where $\tilde{m}_{k} = m_{k} / \langle m \rangle$)

\begin{equation}
  dl^{2} \equiv \sum_k \tilde m _k  \big( d\br^{(k)} \big)^{2}.
\end{equation}
We here derive the discrete \textit{NVU} algorithm applying the Hertzian metric (this appendix also corrects a typo in Eq. (A5) of Paper II). 
The discretized variational condition for geodesic motion on $\Omega$ is

\begin{equation}\label{temp1}
  \delta \left( \sum_i \sqrt{\sum_k \tilde{m}_{k} \left( \textbf{r}_{i}^{(k)} - \textbf{r}_{i-1}^{(k)} \right )^2 } -\sum_i \lambda_i U( \bR_i)\right) \,=\,0\,.
\end{equation}
Assuming a constant step length $l_{0}$, i.e.,

\begin{equation}\label{ansatz1}
  \sum_k \tilde{m}_{k} \left( \textbf{r}_{i}^{(k)} - \textbf{r}_{i-1}^{(k)} \right )^2 \equiv l_{0}^{2},
\end{equation}
it follows by differentiation with respect to $\textbf{r}_{i}^{(k)}$ from Eq. (\ref{temp1}) that

\begin{equation}\label{some1}
  \tilde{m}_{k}\big( \textbf{r}_{i}^{(k)} - \textbf{r}_{i-1}^{(k)} \big) + \tilde{m}_{k}\big( \textbf{r}_{i}^{(k)} - \textbf{r}_{i+1}^{(k)} \big)  + l_{0}\lambda_{i}\textbf{f}_{i}^{(k)} \,=\,0.
\end{equation}
Defining $\textbf{a}^{(k)}_{i} \equiv (\textbf{r}_{i}^{(k)} - \textbf{r}_{i-1}^{(k)})$ and $\textbf{b}^{(k)}_{i} \equiv (\textbf{r}_{i}^{(k)} - \textbf{r}_{i+1}^{(k)})$, Eq. (\ref{ansatz1}) expresses that 
$\sum_{k} \tilde{m}_{k}((\textbf{a}^{(k)}_{i})^{2} - (\textbf{b}^{(k)}_{i})^{2})$ = $\sum_{k} \tilde{m}_{k}(\textbf{a}^{(k)}_{i} + \textbf{b}^{(k)}_{i})(\textbf{a}^{(k)}_{i} - \textbf{b}^{(k)}_{i})$ = 0, 
and thus via Eq. (\ref{some1})

\begin{equation}
  \sum_{k} \tilde{m}_{k}\big(-l_{0}/\tilde{m}_{k}\lambda_{i}\textbf{f}_{i}^{(k)}\big)\big(\textbf{r}^{(k)}_{i+1} - \textbf{r}^{(k)}_{i-1}\big) = 0.
\end{equation}
Equivalently,

\begin{equation}\label{temp2}
  \sum_{k} \textbf{f}_{i}^{(k)}\textbf{r}^{(k)}_{i+1} =  \sum_{k} \textbf{f}_{i}^{(k)} \textbf{r}^{(k)}_{i-1}.
\end{equation}
Combining Eq. (\ref{temp2}) with the discrete \textit{NVU} algorithm (Eq. (\ref{some1})) gives the following result

\begin{equation}\label{lambda}
  l_{0}\lambda_{i} = \frac{-2\sum_{k} \textbf{f}_{i}^{(k)}\cdot(\textbf{r}^{(k)}_{i} - \textbf{r}^{(k)}_{i-1}) }{\sum_{k} \frac{(\textbf{f}_{i}^{(k)})^{2}}{\tilde{m}_{k}}}.
\end{equation}
The atomic \textit{NVU} algorithm with varying masses is thus given by

\begin{align}
  &  \textbf{r}_{i+1}^{(k)} = 2\textbf{r}_{i}^{(k)} - \textbf{r}_{i-1}^{(k)} + \frac{l_{0}}{\tilde{m}_{k}}\lambda_{i}\textbf{f}_{i}^{(k)}, \\
  & l_{0}\lambda_{i} = \frac{-2\sum_{k} \textbf{f}_{i}^{(k)}\cdot(\textbf{r}^{(k)}_{i} - \textbf{r}^{(k)}_{i-1}) }{\sum_{k} \frac{(\textbf{f}_{i}^{(k)})^{2}}{\tilde{m}_{k}}}.
\end{align}


%

\end{document}